\documentclass[prb,aps,twocolumn,superscriptaddress]{revtex4-1}
\usepackage{latexsym}
\usepackage{amssymb}
\usepackage{amsmath}
\usepackage{amsbsy}
\usepackage{mathtools}
\usepackage{mathrsfs}
\usepackage{physics}
\usepackage[pdftex]{graphicx}
\usepackage{epstopdf}
\usepackage[usenames, dvipsnames]{color}
\usepackage{soul}
\usepackage[english]{babel}
\usepackage[normalem]{ulem}

\newcommand{\av}[1]{\bar{#1}}

\newcommand{\blue}[1]{#1}

\newcommand{\unamone}{Departamento de Sistemas Complejos, Instituto de Fisica,
Universidad Nacional Aut\'onoma de M\'exico, Apartado Postal 20-364,01000,
Ciudad de M\'exico, M\'exico.}
\newcommand{\unamtwo}{Instituto de F\'isica,
Universidad Nacional Aut\'onoma de M\'exico, 
Apartado Postal 20-364 01000,Ciudad de  M\'{e}xico, M\'exico}

\newcommand{\insa}{Universit\`e de Toulouse,
INSA-CNRS-UPS, LPCNO, 135 avenue de Rangueil,
31077 Toulouse, France}

\newcommand{\ioffe}{Ioffe Physical-Technical
Institute, 194021 St. Petersburg, Russia}

\newcommand{\uama}{\'Area de F\'isica Te\'orica y
Materia Condensada, Universidad Aut\'onoma Metropolitana
Azcapotzalco, Av. San Pablo 180, Col. Reynosa-Tamaulipas,
02200 Cuidad de M\'exico, M\'exico}

\begin{document}

\title{Electron-nucleus
spin correlation conservation of the spin dependent
recombination in Ga$^{2+}$ centers.}

\author{J. C. Sandoval-Santana}
\affiliation{\unamtwo}
\author{V. G. Ibarra-Sierra}
\affiliation{\unamone}
\author{H. Carr\`ere}
\affiliation{\insa}
\author{L.A. Bakaleinikov}
\author{V. K. Kalevich}
\author{E. L. Ivchenko}
\affiliation{\ioffe}
\author{X. Marie}
\author{T. Amand}
\author{A. Balocchi}
\affiliation{\insa}
\author{A. Kunold}
\affiliation{\uama}


\begin{abstract}
Spin dependent recombination in GaAsN offers many interesting
possibilities in the design of spintronic devices
mostly due to its astounding capability to
reach conduction band electron spin polarizations close to 100\%
at room temperature.
The mechanism behind the spin selective capture of electrons in Ga$^{2+}$
paramagnetic centers is revisited in this paper to address
inconsistencies \blue{common to most} previously presented models.
Primarily, these errors manifest themselves as
major disagreements with the experimental observations of
two key characteristics of this phenomenon:
the effective Overhauser-like magnetic field
and the width of the photoluminescence Lorentzian-like curves
as \blue{a} function of the illumination power.
\blue{These} features are not only essential to understand the spin
dependent recombination in GaAsN, but are also key to the design
of novel spintronic devices.
\blue{Here we} demonstrate that the particular structure of the electron
capture expressions introduces spurious electron-nucleus correlations
that artificially alter the balance between the hyperfine and
the Zeeman contributions. This imbalance strongly distorts
the effective magnetic field and width characteristics.
\blue{In this work} we propose an alternative recombination mechanism that
preserves the electron-nucleus correlations and, at the same
time, keeps the essential properties of the spin selective
capture of electrons.
This mechanism yields a significant improvement
to the agreement between experimental and theoretical results.
In particular, our model gives results in \blue{very good accord
 with the experimental
 effective Overhauser-like magnetic field and
 width data, and with} the degree of circular polarization under
oblique magnetic fields.
\end{abstract}

\maketitle

\section{Introduction}
Spin dependent recombination (SDR) has been studied extensively,
both experimentally
and theoretically because of the many possibilities
it offers to the design of spintronic devices
\cite{PhysRevB.6.436,WEISBUCH1974141,PhysRevB.30.931,
Kalevich2005,doi:10.1063/1.2150252,Kalevich2007,
doi:10.1002/pssa.200673009,Zhao_2009,wang2009room,KALEVICH20094929,
doi:10.1063/1.3186076,doi:10.1063/1.3273393,doi:10.1063/1.3275703,
doi:10.1063/1.3299015,Ivchenko_2010,PhysRevB.83.165202,
PhysRevB.85.035205,doi:10.1063/1.4816970,Kalevich2013,
puttisong2013efficient,PhysRevB.87.125202,PhysRevB.90.115205,
PhysRevB.91.205202,Ivchenko2016,PhysRevB.95.195204,
PhysRevB.97.155201,Sandoval-Santana2018,ibarra2018spin,
chen2018room}.
SDR was first observed in silicon \cite{PhysRevB.6.436,WEISBUCH1974141,PhysRevB.30.931} 
and later \blue{in} GaAsN alloys with a small content of nitrogen \cite{Kalevich2005,Kalevich2007}.
It relies \blue{on the Pauli} principle that states that two electrons \blue{cannot}
occupy simultaneously a quantum level with the same spin orientation.
Hence, the capture rate in a paramagnetic center is strongly
influenced by the relative spin orientation of the
conduction band (CB) electrons and the centers outer shell electrons.
More specifically, the recombination when both electrons
have opposite spin orientations
will have significantly faster capture rates than those with parallel
orientation\cite{doi:10.1063/1.2150252,Kalevich2005,Kalevich2007}.
This difference gives rise to a spin filtering process where
photogenerated electrons are either fastly recombined,
if they have opposite spin orientation to the centers, or
remain for very long times in the \blue{CB if} they have opposing
spin orientations\cite{wang2009room,Zhao_2009,Ivchenko_2010}. 
The spin filtering effect along with the particular selection rules
of GaAs enable to control the degree of CB electron spin polarizations
through optically oriented pumping over a wide energy excitation range\cite{meier2012optical}.
The most staggering outcome of this dynamical process is the
large CB electron spin polarization of almost $100\%$\cite{PhysRevB.91.205202}
that can be attained at room temperature.
Moreover, the presence of large CB electron populations,
sustained by the spin filtering effect, considerably increase the
photoconductivity under the incidence of circularly  polarized light.
This allows \blue{for} the detection of the CB electrons degree of spin polarization
\cite{doi:10.1063/1.3273393,PhysRevB.83.165202}.

\begin{figure*}
\includegraphics[width=0.25\textwidth,keepaspectratio=true]{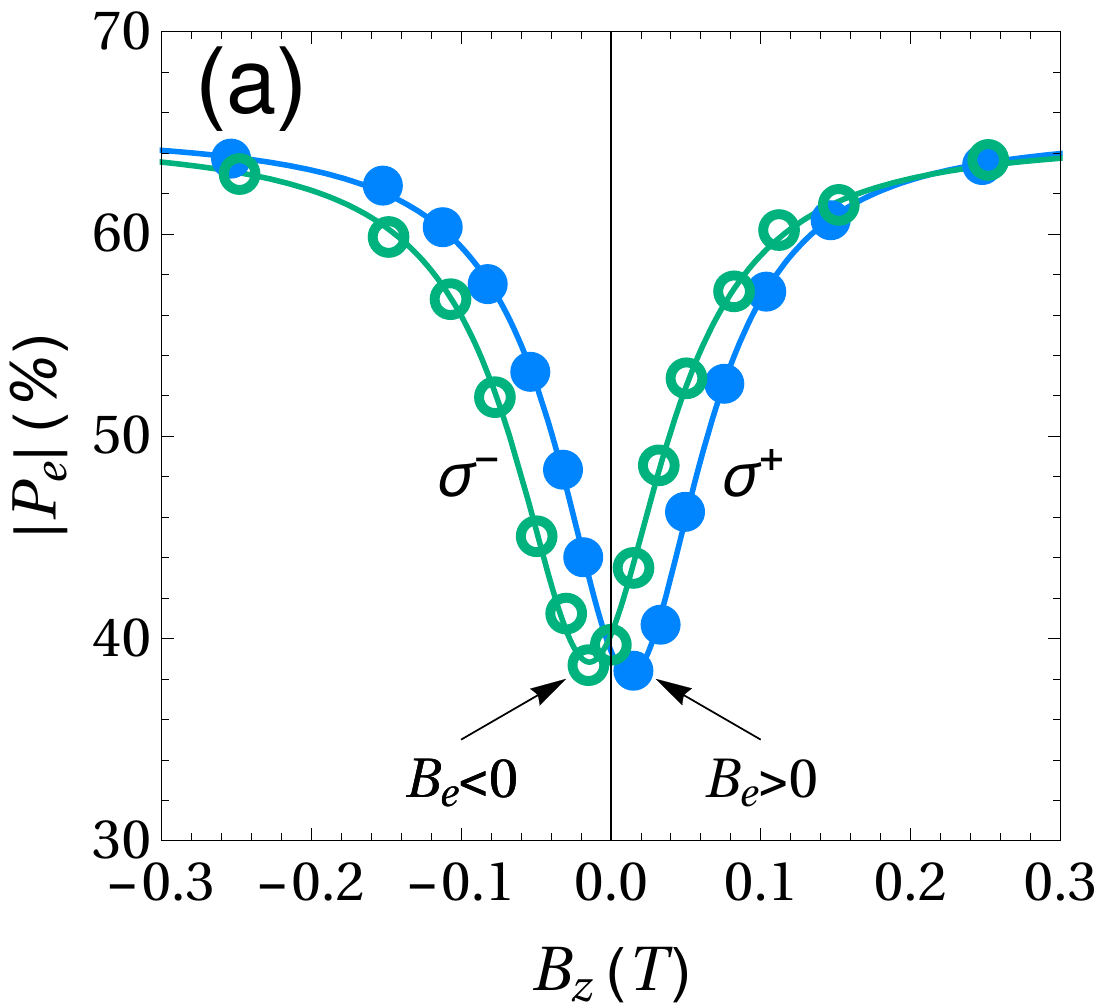}
\includegraphics[width=0.36\textwidth,keepaspectratio=true]{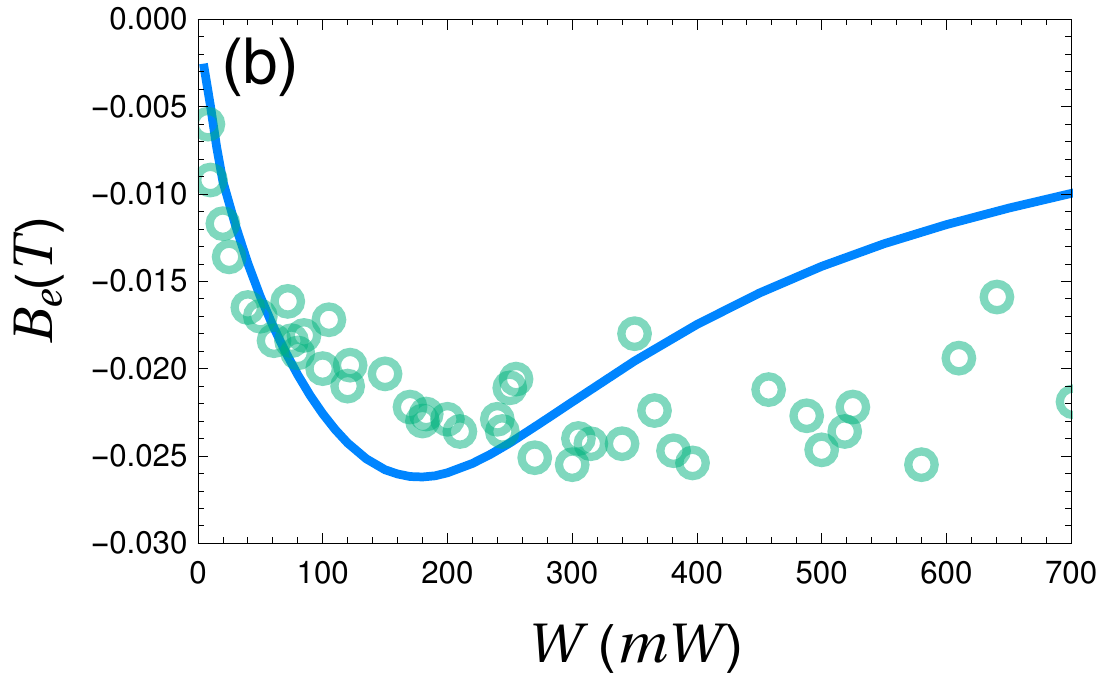}
\includegraphics[width=0.34\textwidth,keepaspectratio=true]{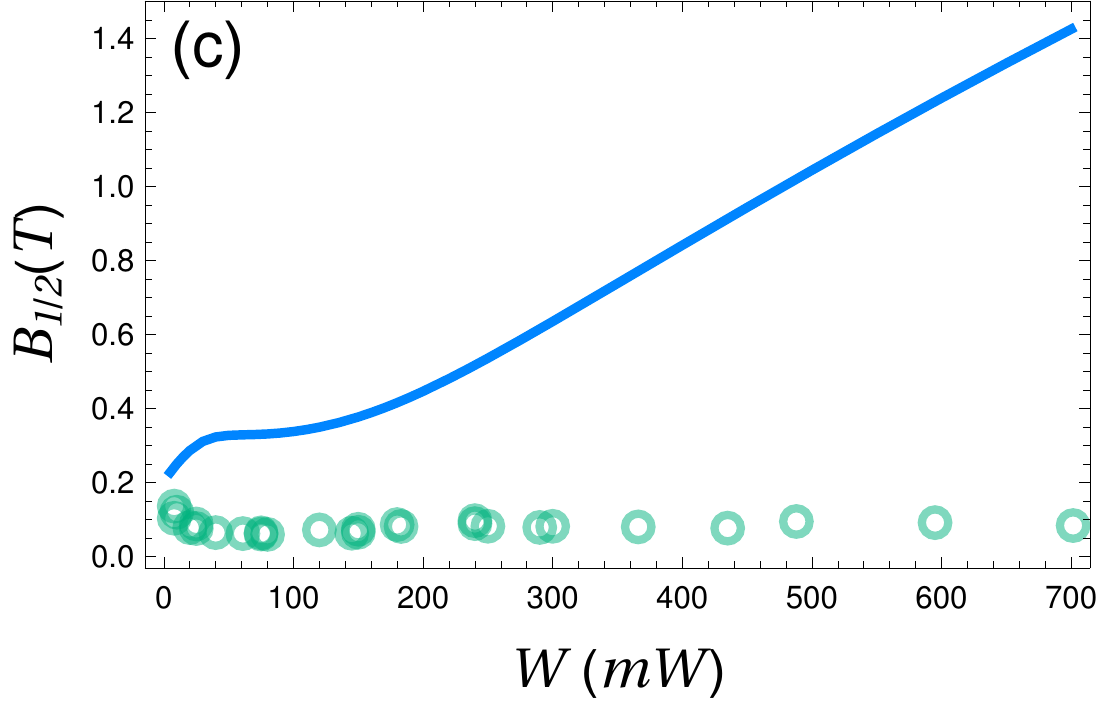}
\caption{(a) Experimental behaviour of the degree of circular
polarization $P_e$ as a function
of the longitudinal magnetic field $B_z$ under right ($\sigma^+$)
and left ($\sigma^-$) circularly polarized light.
(b) Effective magnetic field $B_{\mathrm{e}}$
as a function of the illumination power. 
(c) Mean width $B_{1/2}$ as a function of
the illumination power. The purple solid circles
correspond to the experimental results and the solid blue line
are the theoretical calculations.}
\label{fig:figure1}
\end{figure*}

It is widely accepted that in GaAsN dilute semiconductors,
it is primarily Ga$^{2+}$
interstitial centers that play the role of paramagnetic traps
and spin filtering defects
\cite{doi:10.1063/1.3275703,wang2009room,
doi:10.1063/1.4816970,ibarra2018spin} .
Experimental findings on GaAsN
showed an improvement of the spin filtering effect, as an increase  the photoluminescence (PL) intensity or
the \blue{the degree of circular polarization (DCP)} of the emitted light under a
moderate ($100$mT) Faraday configuration magnetic field
\cite{PhysRevB.85.035205,Kalevich2013,PhysRevB.87.125202,
puttisong2013efficient}.
Specifically, the PL intensity $J(B_z)$ or the DCP $P_e(B_z)$
of the emitted light as a function of the longitudinal magnetic
field $B_z$ take the shape of an inverted Lorentzian \blue{function\sout{s}} as can
be seen in Fig. \ref{fig:figure1} (a).
The primary cause of this phenomenon was identified as
the hyperfine interaction (HFI) between the bounded electrons and the
corresponding nuclei in Ga$^{2+}$ centers
\cite{PhysRevB.87.125202,puttisong2013efficient,PhysRevB.90.115205,
PhysRevB.91.205202,PhysRevB.95.195204}.
The shape of $J(B_z)$ and $P_e(B_z)$ emerges
from the competition between the hyperfine and the Zeeman \blue{interactions}.
In the low magnetic field regime the HFI is dominant, but
as the magnetic field increases and the Zeemna interaction
becomes stronger, bound electrons and nuclei decouple.
At this stage the angular momentum transfer and mixing
between electrons
and nuclei induced by the HFI is interrupted
and the spin filtering effect becomes more
efficient\cite{PhysRevB.90.115205,PhysRevB.95.195204}.
This alone can not fully explain the amplification
of the spin filtering effect.
The nuclear spin relaxation, dominated by the dipolar interaction
between neighbouring nuclei \cite{PhysRevB.95.195204}, also contributes
to the amplification of the spin filtering effect. In fact, in its absence,
the increase in the DCP of the PL
would not be observable \cite{PhysRevB.95.195204}.
The interplay between the HFI and the spin relaxation produces
an effective {Overhauser-like} magnetic field that manifests as a
shift of a few tens of mT of the minimum of $J(B_z)$ and $P_e(B_z)$
with respect to {$B_z=0\,$} mT$\,$\cite{Kalevich2013} as
can be seen in Fig. \ref{fig:figure1} (a).
This shift is a distinctive property of centers 
with nuclear spin $I>1/2$ as \blue{it is the case for Ga$^{2+}$ ($I^{Ga}=3/2$)}
\cite{PhysRevB.30.931,PhysRevB.91.205202,Ivchenko2016}.
Furthermore, the effective magnetic field is sensitive to
the orientation of the circularly polarized light:
the incidence of left circularly polarized light displaces
the curves to the {positive $\rightarrow$ negative } magnetic field region
and right circularly polarized light to the { positive} region.

To fully exploit this properties in spintronic
devices, it is of great importance to completely
comprehend the mechanisms behind the complex
behaviour of Ga$^{2+}$ centers .
Hitherto, experimental results on
PL and DCP under circularly polarized light
in different magnetic field configurations
have been correctly reproduced by four theoretical models
\cite{puttisong2013efficient,PhysRevB.90.115205,
Ivchenko2016,PhysRevB.95.195204}
(a brief summary of them \blue{together with the sample
characteristics and experimental conditions} can be found in
Ref. [\onlinecite{PhysRevB.95.195204}]).
However, not even the most general of these
models\cite{PhysRevB.95.195204} is able
to capture two important features: the effective
magnetic field and the width of the
$J(B_z)$ and $P_e(B_z)$ curves as a function
of the power of the incident light.
Whereas, experimental results yield
a monotonically increasing
effective magnetic field $B_\mathrm{e}$ as a function of the
illumination power until it saturates at approximately
$25$mT \cite{Kalevich2013}, the theoretical results give
a vanishing effective magnetic field in the high power
regime\cite{PhysRevB.95.195204}.
In Fig. \ref{fig:figure1} (b) we have superimposed
the experimental (solid circles) and theoretical (solid line)
behaviours of the effective
magnetic field $B_\mathrm{e}$ as functions of the illumination power.
Likewise, the width $B_{1/2}$ of the $J(B_z)$ and $P_e(B_z)$
experimental curves consist of
monotonically decreasing functions of power that saturate at
approximately $100$mT. The theoretical model gives instead
a monotonically increasing function of power that never saturates.
This can be observed in Fig. \ref{fig:figure1} (c),
where we have plotted the experimental (solid cirlces) and
theoretical (solid line) curves of $B_{1/2}$ vs. power intensity.
Since the two features, the width and the shift
(\blue{the} effective magnetic field),
parametrize the DCP of the emitted light and its orientation,
understanding their origin is a crucial step
towards the design of GaAsN spintronic devices.

In this work we develop a model for the
SDR in Ga$^{2+}$ centers that correctly
accounts for the experimental behaviour
of the width and the shift of the
$J(B_z)$ and $P_e(B_z)$ curves.
It is shown that the recombination
processes considered in previous models
\cite{puttisong2013efficient,
Ivchenko2016,PhysRevB.95.195204}
is incorrect because it artificially introduces
correlations between
bound electrons and nuclei
in G$^{2+}$ centers. The proposed
new SDR mechanism correctly describes
other experiments that are very sensitive to
width and the shift, as the DCP of the
photoluminescence in GaAsN samples
subject to tilted magnetic fields.

The paper is organized as follows.
Section \ref{sec:model} is divided in
several subsections where the master equation
and its multiple elements are introduced.
First we briefly review
the master equation approach adopted to
model the spin dynamics of GaAsN bulk alloy
in section \ref{subsec:master}.
Section \ref{subsec:Liealgebra} deals with
the Lie algebraic method used to build
the dissipators corresponding to
the VB hole and CB electron recombination processes.
These are calculated in sections \ref{subsec:recholes}
and \ref{subsec:sdr}.
In section \ref{subsec:sdr} we address the SDR
process. We show that previous versions
of the SDR dissipator has problems that
introduce artificially electron-nucleus spin correlations.
In this section we propose a new SDR dissipator
that solves this inconsistency.
The comparison of the results produced by the new dissipator
and experimental data
is presented section \ref{sec:resanddisc}.

\section{Model}\label{sec:model}
\subsection{The master equation for GaAsN}\label{subsec:master}
For the sake of completeness,
as a starting point we briefly present the main elements
of the master equation approach used in our
previous works.
The master equation is given by
\begin{equation}
\frac{d\rho}{dt}=\frac{i}{\hbar}\left[\rho,H\right]
  +\mathcal{D}\left(\rho\right),
\label{eq:master}
\end{equation}
where $\rho$ is the density matrix, $H$ is the
Hamiltonian and $\mathcal{D}$ is the dissipator.
We consider the system as being formed
of four subspaces: VB holes, CB electrons,
singly occupied traps and doubly occupied traps.
VB holes are considered to be unpolarized
due to their fast spin relaxation time and
CB electrons have the to possible spin projections of the spin.
Singly occupied traps are Ga$^{2+}$ centers
whose $4s$ shell only has one electron with any
of the two possible projections of spin.
Doubly occupied traps, instead,
are Ga$^{2+}$ centers
whose $4s$ shell if full and therefore cannot capture
any additional electrons due to \blue{the} Pauli principle.
Provided that the coherences between the subspaces
vanish, the density matrix can thus be expressed as
the direct sum
\begin{equation}
\rho=\rho_v \oplus
 \rho_c \oplus
 \rho_1 \oplus
 \rho_2 ,
\end{equation}
where $\rho_{v}$ and $\rho_{c}$ are
valence and conduction band density matrices.
The subspaces of singly and doubly occupied Ga$^{2+}$ centers are
described by the $\rho_1$ and $\rho_2$ density matrices. 
The Hamiltonian for the Zeeman and hyperfine couplings
is given by
\begin{equation}
H=\hbar\boldsymbol{\omega}\cdot\boldsymbol{S}
+\boldsymbol{\Omega}\cdot\boldsymbol{S}
+A\boldsymbol{I}_1\cdot \boldsymbol{S}_c,
\end{equation}
where
$A$ is the hyperfine parameter of
Ga$^{2+}$ centers,
$\boldsymbol{\omega}=g\mu_B\boldsymbol{B}/\hbar$,
$\boldsymbol{\Omega}=g_c\mu_B\boldsymbol{B}/\hbar$,
$\mu_B$ is the Bohr
magneton, $\boldsymbol{B}$ is the external magnetic
field, $g$ is the gyromagnetic factor for
CB electrons and $g_c$
is the gyromagnetic factor for bound electrons in
Ga$^{2+}$ centers. $\boldsymbol{S}$ and
$\boldsymbol{S}_c$ are the CB electrons and
and bound electrons spin operators.
$\boldsymbol{I}_1$ and $\boldsymbol{I}_2$
are the nuclear spin operators of singly
and doubly occupied
Ga$^{2+}$ centers.
The dissipator is given by the sum of the
following contributions
\begin{equation}
\mathcal{D}\left(\rho\right)=\mathcal{G}+\mathcal{D}_S
   +\mathcal{D}_{SC}
   +\mathcal{D}_1+\mathcal{D}_2
   +\mathcal{D}_P+\mathcal{D}_{SDR}.
\end{equation}
The photogeneration of electron-hole pairs
is described by the first term
{
\begin{equation}
\mathcal{G}=\left(G_{+}+G_{-}\right)\left(p+n\right)
+2\left(G_{+}-G_{-}\right)\boldsymbol{e}\cdot \boldsymbol{S},
\end{equation}
}
where $n=2S_0$ is the number operator for CB electrons and
$\boldsymbol{e}$ is a unitary vector
parallel to excitation \blue{direction}.
Spin-up and spin-down CB electron generation rates
are given by the smooth step function
\begin{equation}
G_\pm(t) =\frac{G_0 W}{2}
\frac{1\pm P}{2}\left[
1+\tanh\left(\frac{t-t_0}{s_t}\right)
\right],
\end{equation}
where $G_0$ is the power generation factor,
$W$ is the excitation power
$s_t=10$\,ps is the width of the step function
and $P$ is the spin polarization
degree ($0<P\le 1$ generates mostly spin-up electrons
and $-1\le P< 0$ spin-down electrons).
The CB and bound electron spin relaxation
are given by
\begin{eqnarray}
\mathcal{D}_S&=&-\frac{1}{2\tau_s}\sum_{k=1}^{3}\left[S_k,\left[S_k,\rho\right]
\right],\\
\mathcal{D}_{SC}&=&-\frac{1}{2\tau_{sc}}
\sum_{k=1}^{3}\left[S_{ck},\left[S_{ck},\rho\right]
\right].
\end{eqnarray}
Assuming that the nuclear spin relaxation
is dominated by dipole-dipole interaction
between neighbouring nuclei,
Wangsness-Bloch-Redfield theory
states that the dissipators take the form
\cite{PhysRev.89.728,REDFIELD19651,
PhysRev.142.179,PhysRevB.95.195204}
\begin{eqnarray}
\mathcal{D}_1&=&-\frac{1}{3\tau_{n1}}
\sum_{k=1}^{3}\left[I_{1k},\left[I_{1k},\rho\right]
\right],\\
\mathcal{D}_2&=&-\frac{1}{3\tau_{n2}}
\sum_{k=1}^{3}\left[I_{2k},\left[I_{2k},\rho\right]
\right].\label{nucdis}
\end{eqnarray}

\subsection{Lie algebraic approach}\label{subsec:Liealgebra}
The recombination of VB holes and
CB electrons require a special
mathematical treatment.
The process of holes recombining into doubly occupied
traps and leaving a singly occupied one
is represented by the $\mathcal{D}_P$ dissipator.
The SDR dissipator $\mathcal{D}_{SDR}$,
the central subject of our discussion,
accounts for the
spin dependent capture of a CB electron into a
a singly occupied trap creating a doubly occupied one.
The dissipators $\mathcal{D}_P$ and $\mathcal{D}_{SDR}$
can be conveniently expanded as the superposition
of the elements of an internal space of Hermitian
matrices that span the space of the four subspaces.
This way of proceeding
has many advantages. First, it
easily helps avoid including
spurious coherences, reducing the number
of unknowns. Second, the internal product
allows expressing
any operator in a simple way
and enables working out the dynamical equations of the system.
And third, it considerably simplifies building
both dissipators.
The internal space is constituted of the following elements 
\begin{multline}
\Lambda=\left\{p,S_k,U_{k,j,i},V_{j,i},
\right\}\\
=\left\{\lambda_1,\lambda_2,\dots,\lambda_{d}\right\}
,\,\,\,\, i,j,k=0,1,2,3 \,\,\, ,\label{base}
\end{multline}
where $d=85$ is the dimension of the algebra.
The previous operators are explicitly given by
\begin{eqnarray}
p&=&1_{1\times 1}\oplus 
  0_{2\times 2}
  \oplus 0_{8\times 8} 
  \oplus 0_{4\times 4},
\\
S_k&=& 0_{1\times 1}
  \oplus \left(s_k\right)
  \oplus 0_{8\times 8} 
  \oplus 0_{4\times 4},
\\
U_{k,j,i} &=& 0_{1\times 1}
  \oplus 0_{2\times 2}\oplus
  \left(s_k\otimes s_j\otimes s_i\right)
  \oplus 0_{4\times 4},
\\
V_{j,i} &=& 0_{1\times 1} 
  \oplus 0_{2\times 2}
  \oplus 0_{8\times 8}
  \oplus\left(s_j\otimes s_i\right),
\end{eqnarray}
where $i,j,k=0,1,2,3$.
In this notation $k$ is related to
the electron spin and $i$ and $j$ to
the nuclear spin.
The $s_k$ spin matrices
follow the standard commutation rules
\begin{equation}
\left[s_i,s_j\right]=i\hbar \sum_k\epsilon_{i,j,k}s_k, \,\,\,\, i,j,k=1,2,3,
\end{equation}
and $s_0=(1/2)1_{2\times 2}$ is half the identity matrix.
The population of VB holes is represented by the operator $p$.
The population of CB electrons is $n=2S_0$ and
$S_k$ with $k=1,2,3$ are the spin components of CB electrons.
The spin matrices for Ga$^{2+}$ centers
are given by
\begin{equation}
S_{ck} = 4U_{k,0,0},\,\,\, k=1,2,3 \,\,.
\end{equation}
The population operators of singly and doubly
occupied centers are
\begin{eqnarray}
N^1 &=& 8U_{0,0,0} \,\,\, ,\\
N^2 &=& 4V_{0,0} \,\,\, .
\end{eqnarray}

The elements of $\Lambda$ form a Lie algebra and
are the orthogonal
elements of an inner product vector space whose inner product is
given by the trace
\begin{equation}
\left(\lambda_q,\lambda_{q^{\prime}}\right)=
\tr\left[\lambda_q^{\dag}\lambda_{q^{\prime}}\right]=\Tr\left[\lambda_q^2\right]
\delta_{q,q^{\prime}}.\label{internal}
\end{equation}
Even though one can in principle
define many other different inner products
for the elements of $\Lambda$, this one has the
additional advantage of being closely linked to the
quantum statistical average
\begin{equation}
    \av{O}=\Tr\left[O\rho \right],
    \label{eq:expectation}
\end{equation}
where $O$ is any given operator in the same space as $\rho$
and the upper bar indicates the quantum statistical average.
This equation implies that the density matrix
can be expanded in terms of the quantum statistical averages
of the elements of $\Lambda$ as
\begin{equation}
\rho=\sum_q\frac{1}{\Tr\left[\lambda_q^2\right]}\lambda_q\av{\lambda_q}.
\label{eq:densmat}
\end{equation}
Any operator can be likewise expanded in terms
of the elements of $\Lambda$ as
\begin{equation}
O=\sum_q\frac{\Tr[O\lambda_q]}{\Tr [\lambda_q^2]}\lambda_q.\label{eq:expand}
\end{equation}
For example,
the nuclear spin operators can be written as
\begin{eqnarray}
I_{1,k} &=& \sum_{j,i=0}^3 
  \frac{ \Tr\left[I_{1,k}U_{0,j,i}\right]}
  {\Tr\left[U_{0,j,i}U_{0,j,i}\right]}U_{0,j,i}
  \nonumber \\
  &&=8\sum_{j,i=0}^3 
  \Tr\left[I_{1,k}U_{0,j,i}\right]U_{0,j,i},\label{eq:nucspin1}\\
I_{2,k} &=& \sum_{j,i=0}^3 
  \frac{ \Tr\left[I_{2,k}V_{j,i}\right]}
  {\Tr\left[V_{j,i}V_{j,i}\right]}V_{j,i}
  \nonumber \\
  &&=4\sum_{j,i=0}^3 
  \Tr\left[I_{2,k}V_{j,i}\right]V_{j,i}.\label{eq:nucspin2}
\end{eqnarray}

\subsection{Recombination of valence band holes
in paramagnetic centers}\label{subsec:recholes}
Here we build the dissipator $\mathcal{D}_{P}$ for the
process in which holes recombine into Ga$^{2+}$ centers.
This calculation will serve to illustrate
the more complex calculation of $\mathcal{D}_{SDR}$.
The strategy consists in projecting the
well known  two-charge-state kinetic equations\cite{Ivchenko_2010}
onto the base $\Lambda$ in order to translate them
into the dissipator $\mathcal{D}_{P}$.
We start from the rate equations of the two-charge-state
model\cite{Ivchenko_2010}
\begin{eqnarray}
\frac{d}{dt}\av{p} &=& 
  -c_p\, \av{p} \sum_{\beta=-3/2}^{3/2}\av{N}^2_{\beta},\label{eq:ratehole1}\\
\frac{d}{dt}\av{N}^1_{\alpha,\beta} &=& 
  \frac{1}{2}c_p \,\av{p} \av{N}^2_{\beta},\\
\frac{d}{dt}\av{N}^2_{\beta} &=&
  -c_p \,\av{p}\, \av{N}^2_{\beta},\label{eq:ratehole3}
\end{eqnarray}
where $\av{p}=\Tr[p \rho]$ is the quantum statistical
average population of VB holes.
The capture coefficient for holes is $c_p=1/N_0\tau_h$ where
$N_0$ is the total number of centers in the sample
and $\tau_h$ is the hole recombination time in the
high excitation power regime.
The populations of singly and doubly occupied centers
$\av{N}^1_{\alpha,\beta}$ and $\av{N}^2_{\beta}$
are associated with the population operators
$N^1_{\alpha,\beta}$ and $N^2_{\beta}$.
The subscripts $\alpha=-1/2,1/2$ and $\beta=-3/2,-1/2,1/2,3/2$
tag the bound electron and nuclear spin
states $|\alpha,\beta\rangle$ and $|\beta\rangle$
in singly and double charged centers, respectively.
Note that whereas doubly charge states
are indexed by the
electron spin subscript $\alpha$, singly charged centers
are not, because in the latter, both electrons
form a singlet state, rendering the electron spin
index irrelevant.
The singly occupied population operator is thus given by
\begin{equation}
N^1_{\alpha,\beta}=\mathrm{diag}(
  \overbrace{0}^p,
  \overbrace{0,0}^{S_k},
  \overbrace{0,\dots,0,1,0,\dots,0}^{U_{k,j,i}}
  \overbrace{0,0,0,0}^{V_{j,i}}),
\end{equation}
where the $1$
is in the entry corresponding to the state $|\alpha,\beta\rangle$.
Similarly, the doubly occupied population operator is
\begin{equation}
    N^2_{\beta}=\mathrm{diag}(
\overbrace{0}^p,
\overbrace{0,0}^{S_k},
\overbrace{0,0,0,0,0,0,0,0}^{U_{k,j,i}}
\overbrace{0,\dots,1,\dots,0}^{V_{j,i}}),
\end{equation}
where the $1$ is located in
the entry associated to the state $|\beta\rangle$.

Whilst the rate equations (\ref{eq:ratehole1})-(\ref{eq:ratehole3})
merely deal with the density matrix populations,
the elements of $\Lambda$ are of a more general nature
and also involve off-diagonal entries of the density matrix.
Hence, merely projecting the rate equations
onto $\Lambda$ does not suffice to get the most general
form of the dissipator; it is further necessary to demand
that the projected equations comply with basic requirements
as isotropy of space, spin conservation and the usual
tensor transformation rules.
Following the procedure above,
we obtain the set of kinetic equations 
\begin{eqnarray}
\frac{d}{dt}\av{p} &=& 
  -4 c_p\, \av{p}\,\av{V}_{0,0},\label{eq:genrate1}\\
\frac{d}{dt}\av{U}_{k,j,i} &=& 
  \frac{1}{2}\delta_{k,0} c_p\, \av{p}\,\av{U}_{0,j,i},\\
  \frac{d}{dt}\av{V}_{j,i} &=& 
  -c_p\, \av{p}\,\av{V}_{j,i}.\label{eq:genrate3}
\end{eqnarray}
By projecting the right hand side of (\ref{eq:genrate1})-(\ref{eq:genrate3})
onto the elements of $\Lambda$ via Eq. (\ref{eq:expand}),
we get the explicit form of the dissipator
\begin{multline}
\mathcal{D}_P= -\left(4 c_p\, \av{p}\,\av{V}_{0,0} \right)p
  +\frac{1}{8}\left(\frac{1}{2}
  \sum_{j,i=0}^3\av{p}\,\av{U}_{0,j,i}\right)U_{0,j,i}\\
-\frac{1}{4}\left(\sum_{j,i=0}^3\av{p}\,\av{V}_{j,i}\right)V_{j,i}.
\end{multline}
This dissipator generates the rate equations
(\ref{eq:genrate1})-(\ref{eq:genrate3})
when plugged into the master equation (\ref{eq:master}).

\subsection{Spin dependent recombination of
conduction band electrons}\label{subsec:sdr}
Now we turn to the discussion of the spin
dependent recombination dissipator $\mathcal{D}_{SDR}$.
We analyze two different sets of
rate equations. The first one was
presented in Refs. [\onlinecite{puttisong2013efficient}] and
[\onlinecite{PhysRevB.95.195204}].
Even though \blue{these rate equations allow}
to reproduce many of the features of the spin dynamics
in GaAsN, they fail to replicate the width and
the effective magnetic field as functions of the illumination
power. Here we show that, due to
their structure, they artificially alter correlations
between the bound electrons and nuclei in Ga$^{2+}$ centers
during the recombination process.
These rate equation are
\begin{eqnarray}
\frac{d}{dt}\av{n}_{\alpha} &=& 
  -c_n \av{n}_{\alpha}\sum_{\beta=-3/2}^{3/2}\av{N}^1_{-\alpha,\beta},
  \label{eq:ratesdr1}\\
\frac{d}{dt}\av{N}^1_{\alpha,\beta} &=&
  -c_n \av{n}_{-\alpha}\av{N}^1_{\alpha,\beta},
  \label{eq:ratesdr2}\\
\frac{d}{dt}\av{N}^2_{\beta} &=&
  -c_n \sum_{\alpha=-1/2}^{1/2}\av{n}_{-\alpha}\av{N}^1_{\alpha,\beta},
  \label{eq:ratesdr3}
\end{eqnarray}
where $\av{n}_{\alpha}=\Tr[n_\alpha\rho]$
is the quantum statistical population average
of CB electrons with spin
equal to $\alpha$.
The corresponding population operators are given by
\begin{eqnarray}
n_{-1/2} &=& \mathrm{diag}(
  \overbrace{0}^p,
  \overbrace{1,0}^{S_k},
  \overbrace{0,0,0,0,0,0,0,0}^{U_{k,j,i}}
  \overbrace{0,0,0,0}^{V_{j,i}}),\nonumber \\ \\
n_{1/2} &=& \mathrm{diag}(
  \overbrace{0}^p,
  \overbrace{0,1}^{S_k},
  \overbrace{0,0,0,0,0,0,0,0}^{U_{k,j,i}}
  \overbrace{0,0,0,0}^{V_{j,i}}),\nonumber \\
\end{eqnarray}
In the rate $-c_n\av{n}_{-\alpha}\av{N}^1_{\alpha,\beta}$,
the opposite signs of the spin
subscript promote the recombination of CB electrons
onto traps whose bound electrons
have oppositely oriented spin.
The capture coefficient for CB electrons is $c_n=1/N_0\tau^*$ where
$\tau^*$ is the electron recombination time in the
low excitation power regime.

Following the same procedure as in the previous section,
we project these equation on the base $\Lambda$.
The resulting rate equations read
\begin{eqnarray}
\frac{d}{dt}\av{S}_k &=& -4c_n\sum_{k^{\prime},k^{\prime\prime}=0}^3
  \av{S}_{k^{\prime}}
  Q^{\top}_{k,k^{\prime},k^{\prime\prime}}
  \av{U}_{k^{\prime\prime},j,i}\, ,
 \label{eq:ratesdrtens1}\\
\frac{d}{dt}\av{U}_{k,j,i} &=&
 -c_n\sum_{k^{\prime},k^{\prime\prime}=0}^3
  \av{S}_{k^{\prime}}
  Q_{k,k^{\prime},k^{\prime\prime}}
  \av{U}_{k^{\prime\prime},j,i}\, ,
  \label{eq:ratesdrtens2}\\
\frac{d}{dt}\av{V}_{j,i} &=&
  2c_n\sum_{k^{\prime},k^{\prime\prime}=0}^3
  \av{S}_{k^{\prime}}
  Q^{\top}_{0,k^{\prime},k^{\prime\prime}}
  \av{U}_{k^{\prime\prime},j,i}\, ,
  \label{eq:ratesdrtens3}
\end{eqnarray}
and yield the dissipator
\begin{multline}
\mathcal{D}_{SDR}=
  -2\sum_{k=0}^3
  \left(4c_n\sum_{k^{\prime},k^{\prime\prime}=0}^3
  \av{S}_{k^{\prime}}
  Q^{\top}_{k,k^{\prime},k^{\prime\prime}}
  \av{U}_{k^{\prime\prime},j,i}\right)
  S_{k}\\
 -8\sum_{k,j,i=0}^3
 \left(c_n\sum_{k^{\prime},k^{\prime\prime}=0}^3
  \av{S}_{k^{\prime}}
  Q_{k,k^{\prime},k^{\prime\prime}}
  \av{U}_{k^{\prime\prime},j,i}\right)
  U_{k,j,i}\\
 +4\sum_{j,i=0}^3
  \left(2c_n\sum_{k^{\prime},k^{\prime\prime}=0}^3
  \av{S}_{k^{\prime}}
  Q^{\top}_{0,k^{\prime},k^{\prime\prime}}
  \av{U}_{k^{\prime\prime},j,i}\right)
  V_{j,i},\label{eq:dissdr}
\end{multline}
where the matrices $Q_{k,k^{\prime},k^{\prime\prime}}
=\left(\mathbb{Q}_k\right)_{k^{\prime},k^{\prime\prime}}$
are given by
\begin{align}
&\mathbb{Q}_0 =    \left(
\begin{array}{cccc}
 1 & 0 & 0 & 0 \\
 0 & -1 & 0 & 0 \\
 0 & 0 & -1 & 0 \\
 0 & 0 & 0 & -1 \\
\end{array}
\right),&
\mathbb{Q}_1 = \left(
\begin{array}{cccc}
 0 & 1 & 0 & 0 \\
 -1 & 0 & 0 & 0 \\
 0 & 0 & 0 & 0 \\
 0 & 0 & 0 & 0 \\
\end{array}
\right),& \\
& \mathbb{Q}_2 =\left(
\begin{array}{cccc}
 0 & 0 & 1 & 0 \\
 0 & 0 & 0 & 0 \\
 -1 & 0 & 0 & 0 \\
 0 & 0 & 0 & 0 \\
\end{array}
\right), &
\mathbb{Q}_3 =\left(
\begin{array}{cccc}
 0 & 0 & 0 & 1 \\
 0 & 0 & 0 & 0 \\
 0 & 0 & 0 & 0 \\
 -1 & 0 & 0 & 0 \\
\end{array}
\right).
\end{align}
They have the property of
transforming as a scalar for $k=0$ and as a vector for $k\ne 0$.
That is,
\begin{eqnarray}
\mathbb{R}\mathbb{Q}_0\mathbb{R}^\top &=& \mathbb{Q}_0 \, ,\\
\mathbb{R}\mathbb{Q}_k\mathbb{R}^\top &=&
  \sum_{k^{\prime}=1}^3R_{k,k^{\prime}}\mathbb{Q}_{k^{\prime}} \, ,
\end{eqnarray}
where $R_{k,k^{\prime}}=(\mathbb{R})_{k,k^{\prime}}$
is a rotation matrix.
Just as the opposite signs of the electron spins
in Eqs. (\ref{eq:ratesdr1})-(\ref{eq:ratesdr3}),
the $\mathbb{Q}_k$ matrices
are responsible for the spin selective capture of electrons.
Note that while the electron spin indices $k$,$k^{\prime}$ and
$k^{\prime\prime}$ are contracted through the matrix $\mathbb{Q}$,
the ones related with the nuclear spin indices ($j$ and $i$)
are not. This is an indication that, whereas the CB and
bound electron spins must have opposite orientations to recombine, nuclear
spin must be preserved during a capture process.
A further important property of the chosen rate equations
(\ref{eq:ratesdr1})-(\ref{eq:ratesdr3})
and consequently of (\ref{eq:ratesdrtens1})-(\ref{eq:ratesdrtens3})
is that they insure the positive definitness of
the density matrix provided that it has appropriate
initial conditions. This is because the
recombination rate is proportional to both $n_{\alpha}$
and $N^{1}_{-\alpha,\beta}$ hence preventing
any of these populations to reach values below zero.

\begin{figure*}
\includegraphics[width=0.45\textwidth,keepaspectratio=true]{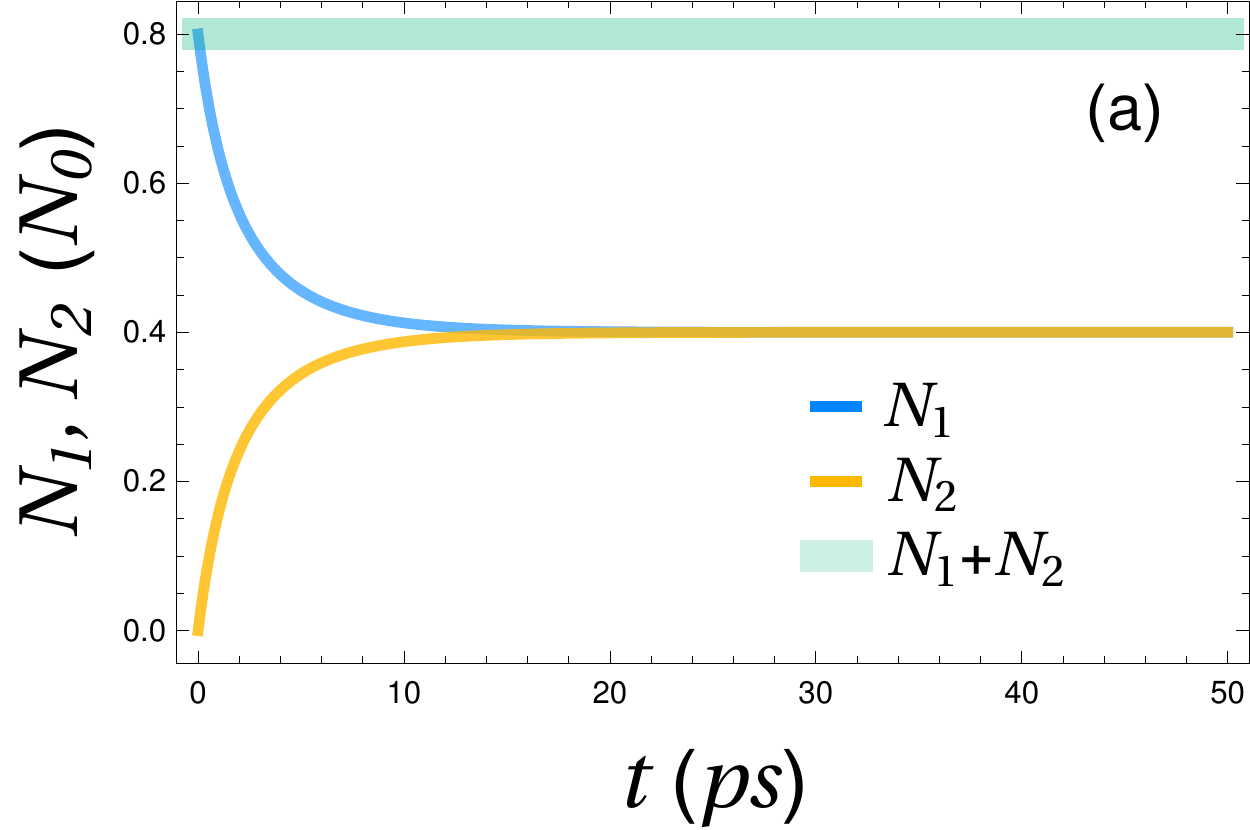}
\includegraphics[width=0.45\textwidth,keepaspectratio=true]{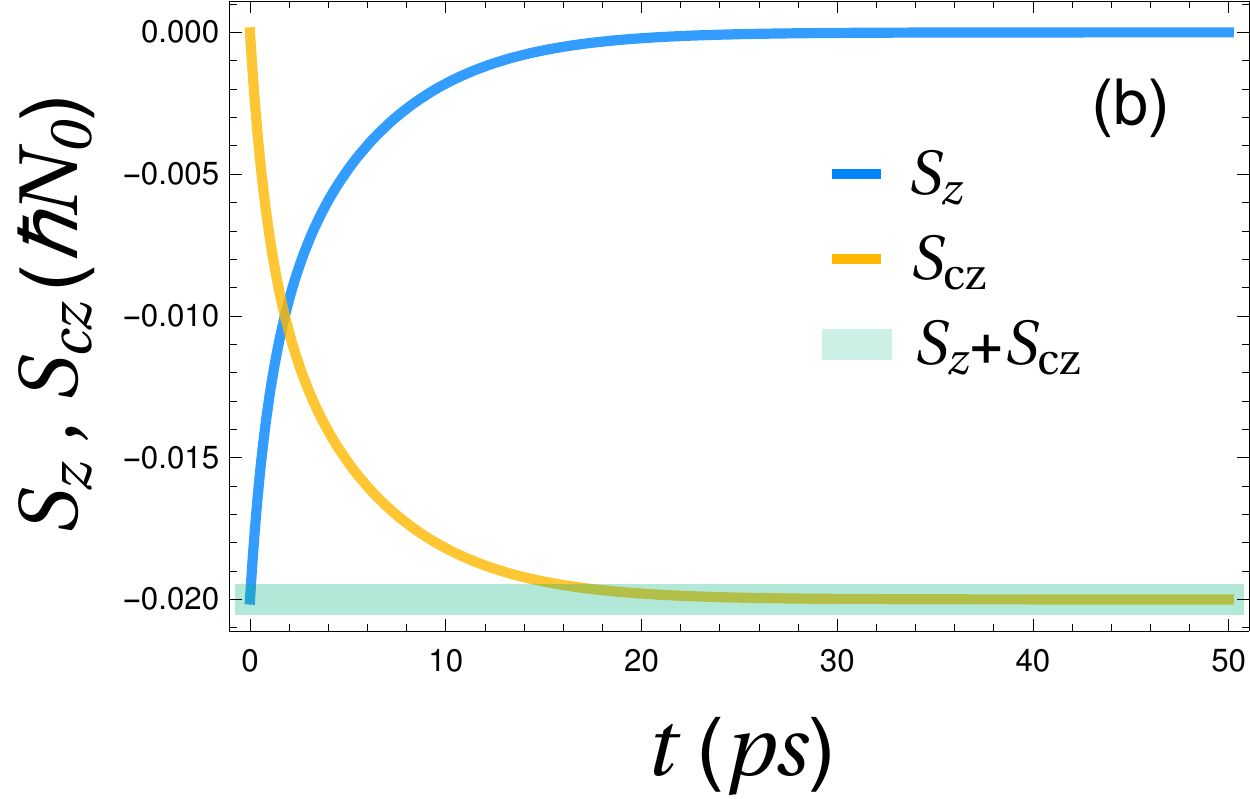}
\includegraphics[width=0.45\textwidth,keepaspectratio=true]{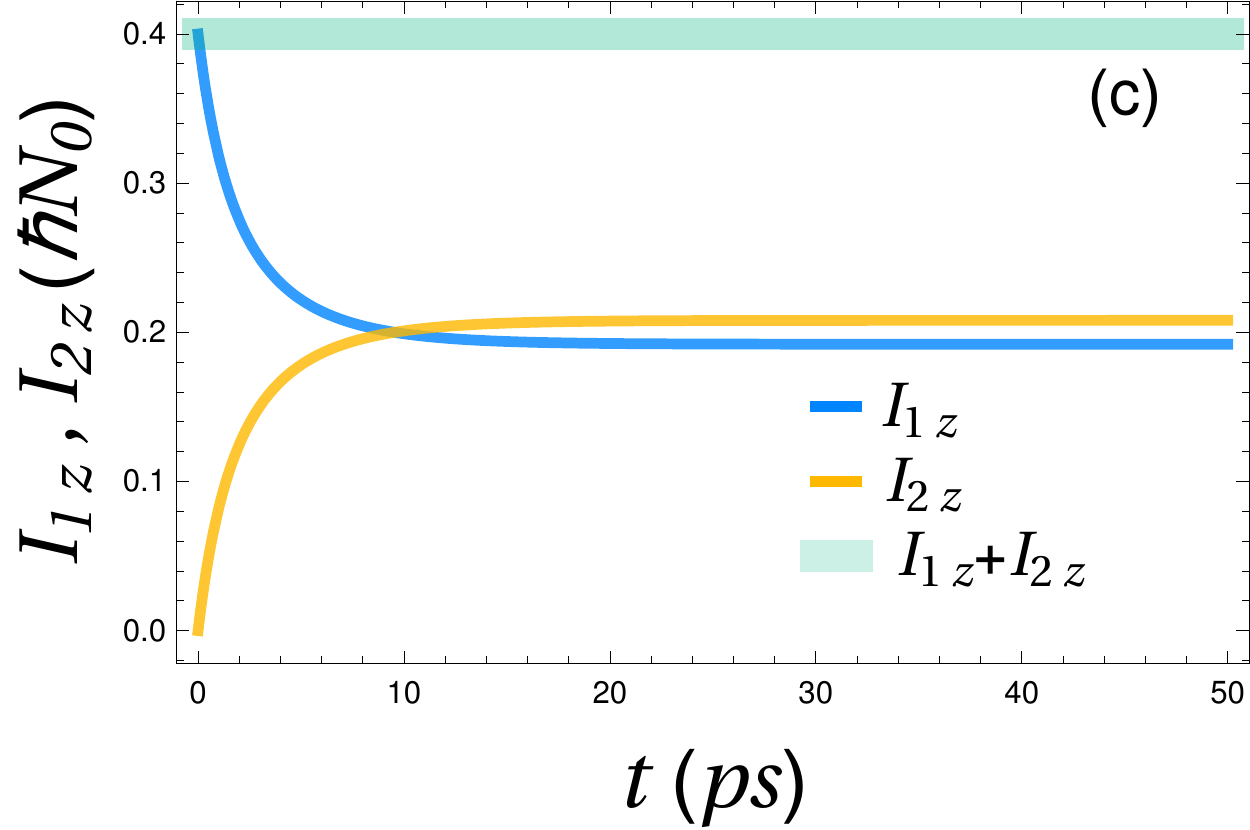}
\includegraphics[width=0.45\textwidth,keepaspectratio=true]{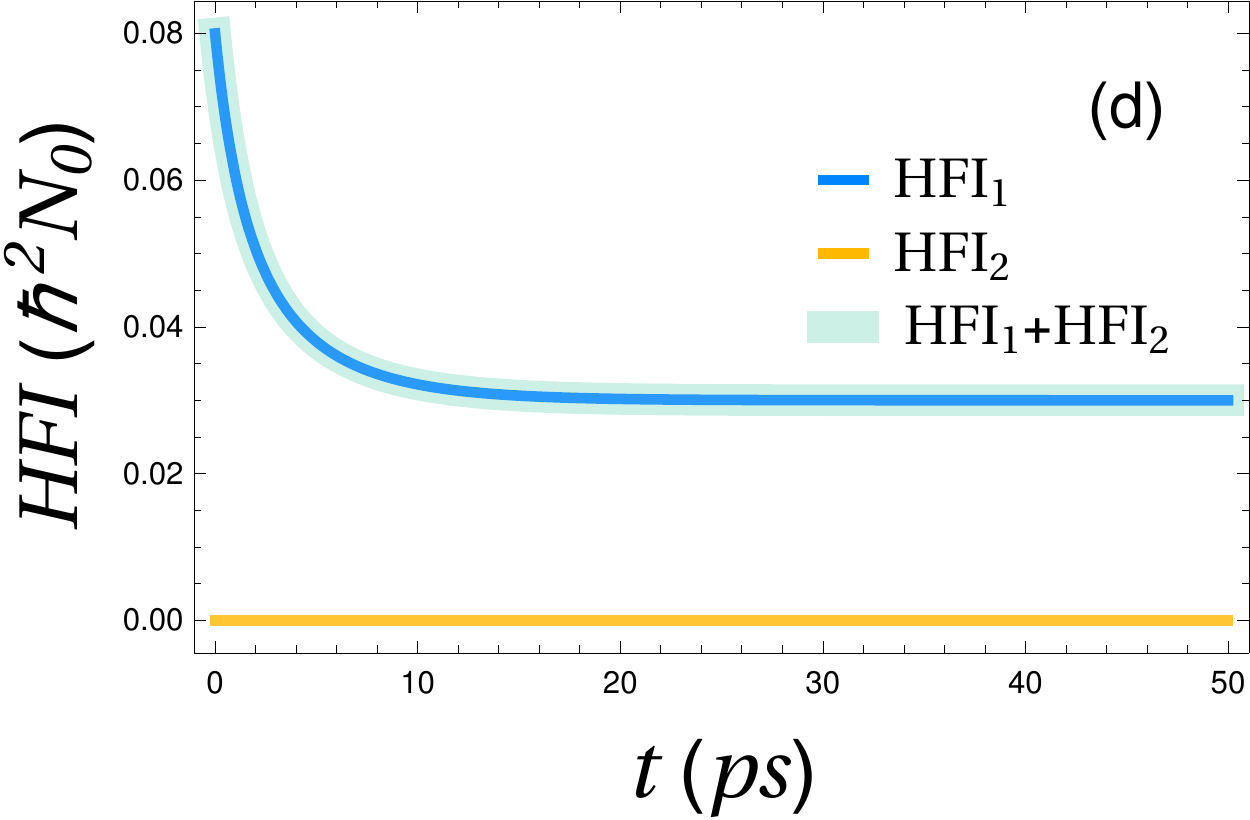}
\caption{(a) Number of singly and doubly occupied centers $N_1$
and $N_2$ as a function of time.
(b) Electronic spin of singly and doubly occupied centers $S_z$
and $S_{cz}$ as a function of time.
(c) Nuclear spin of singly and doubly occupied centers $I_{1z}$
and $I_{2z}$ as a function of time.
(d) Hyperfine interaction as a function of time for
singly and doubly occupied centers. The blue and orange solid lines
correspond to singly and doubly occupied centers respectively.
The thick green line is the sum of singly and doubly occupied centers.
The calculation was performed only taking into account
the SDR dissipator of Eq. (\ref{eq:dissdr}) for the following
initial populations
$n_{-1/2}=0.22 N_0$, $n_{-1/2}=0.18 N_0$, $N_{-1/2,-1/2}$
$=N_{-1/2,1/2}$  $=N_{1/2,3/2}=0.2$,
$N_{-1/2,-3/2}=0.08$, $N_{-1/2,3/2}=0.1$, $N_{1/2,-3/2}=0.02$,
$=N_{1/2,1/2}=0.1$.}
\label{fig:figure2}
\end{figure*}

Let us investigate the impact of this dissipator in the conservation
of populations, electronic spins and nuclear spins.
To do so, we strip off any terms but $\mathcal{D}_{SDR}$ of the master
equation (\ref{eq:master}) .
Charge conservation can be directly proved
by counting the overall number of negatively charged electrons,
either in the CB or in the doubly charged centers,
and positively charged VB holes
\begin{multline}
\frac{d}{dt}\left(\av{p}-\av{n}-\av{N}^2\right)
  =\frac{d}{dt}\left(\av{p}-2\av{S}_0-4\av{V}_{0,0}\right)\\
  =\Tr\left[\left(p-2S_0-4V_{00}\right)\mathcal{D}_{SDR}\right]=0,
  \label{eq:conscharge}
\end{multline}
where in the last term of the right hand side we have substituted
the explicit form of $\mathcal{D}_{SDR}$.
In a similar manner we can demonstrate that $\mathcal{D}_{SDR}$
maintains a constant number of centers
\begin{multline}
\frac{d}{dt}\left(\av{N}^1+\av{N}^2\right)
  =\frac{d}{dt}\left(8\av{U}_{0,0,0}+4\av{V}_{0,0}\right)\\
  =\Tr\left[\left(8U_{0,0,0}+4V_{0,0}\right)\mathcal{D}_{SDR}\right]=0.
  \label{eq:constcenters}
\end{multline}
Electronic spin conservation also holds
\begin{multline}
\frac{d}{dt}\left(\av{S}_k+\av{S}_{ck}\right)
  =\frac{d}{dt}\left(\av{S}_k+4\av{U}_{k,0,0}\right)\\
  =\Tr\left[\left(S_k+4U_{k,0,0}\right)\mathcal{D}_{SDR}\right]=0.
  \label{eq:conselspin}
\end{multline}
Nuclear spin is preserved as well because,
substituting
\begin{multline}
\frac{d}{dt}\left(2\av{U}_{0,j,i}+\av{V}_{j,i}\right)
  =\Tr\left[\left(2U_{0,j,i}+V_{j,i}\right)\mathcal{D}_{SDR}\right]=0,
  \label{eq:nucspinstrucons}
\end{multline}
in $ d/dt((\av{I}_{1,k}+\av{I}_{2,k})
=\Tr[(I_{1,k}+I_{2,k})\mathcal{D}_{SDR}]$,
using Eqs. (\ref{eq:nucspin1})-(\ref{eq:nucspin2})
and $2\Tr[I_{1,k}U_{0,j,i}]=\Tr[I_{2,k}V_{j,i}]$ we obtain
\begin{equation}
    \frac{d}{dt}\left(\av{I}_{1,k}+\av{I}_{2,k}\right)=0.
    \label{eq:nucspincons}
\end{equation}
Eq. (\ref{eq:nucspinstrucons}), however, is of a more general
character than the simple nuclear spin conservation. While
(\ref{eq:nucspincons}) implies the conservation of only
three quantities, (\ref{eq:nucspinstrucons}) implies the
conservation of sixteen.
These quantities, which have the form $2\av{U}_{0,j,i}+\av{V}_{j,i}$,
correspond to those elements
of the density matrix that encode the spin nuclear
structure.
It could be said then that Eq.
(\ref{eq:nucspinstrucons}) means that the
entire nuclear structure, is preserved during the
spin selective recombination of an electron.
Figure \ref{fig:figure2} shows center, electronic spin
and nuclear spin conservation. All four panels
show a situation where initial spin polarized CB
electrons ($n_{-1/2}=0.22 N_0$, $n_{-1/2}=0.18 N_0$)
recombine into an ensemble of singly occupied
centers with inhomogeneously populated nuclear states
($N_{-1/2,-1/2}$ $=N_{-1/2,1/2}$  $=N_{1/2,3/2}=0.2$,
$N_{-1/2,-3/2}=0.08$, $N_{-1/2,3/2}=0.1$, $N_{1/2,-3/2}=0.02$,
$=N_{1/2,1/2}=0.1$).
The solid blue line and the solid orange line correspond
to singly and doubly occupied centers.
The thick green line represents the sum of
singly and doubly occupied centers.
We observe that the number of centers, the electronic spin
and the nuclear spin are conserved
in Figs. (\ref{fig:figure2}) (a), (b) and (c) respectively.

\begin{figure*}
\includegraphics[width=0.45\textwidth,keepaspectratio=true]{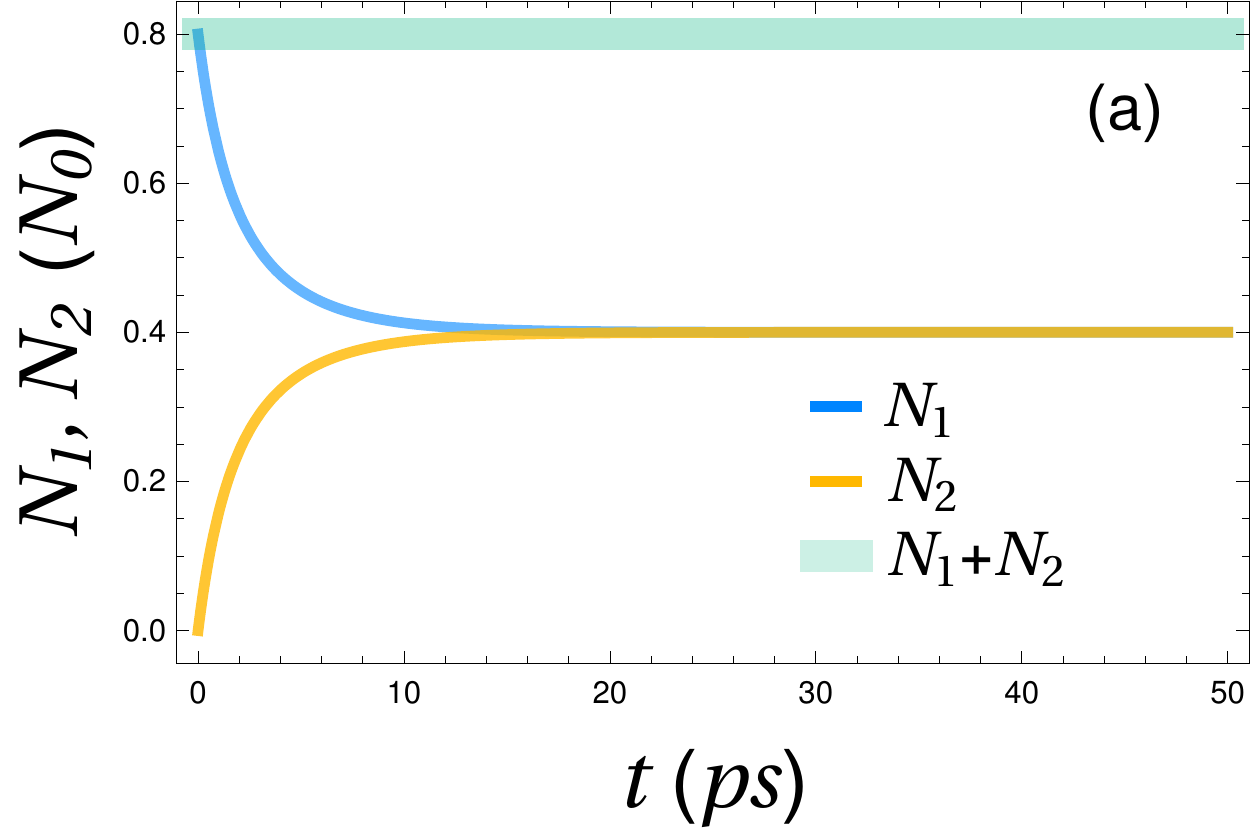}
\includegraphics[width=0.45\textwidth,keepaspectratio=true]{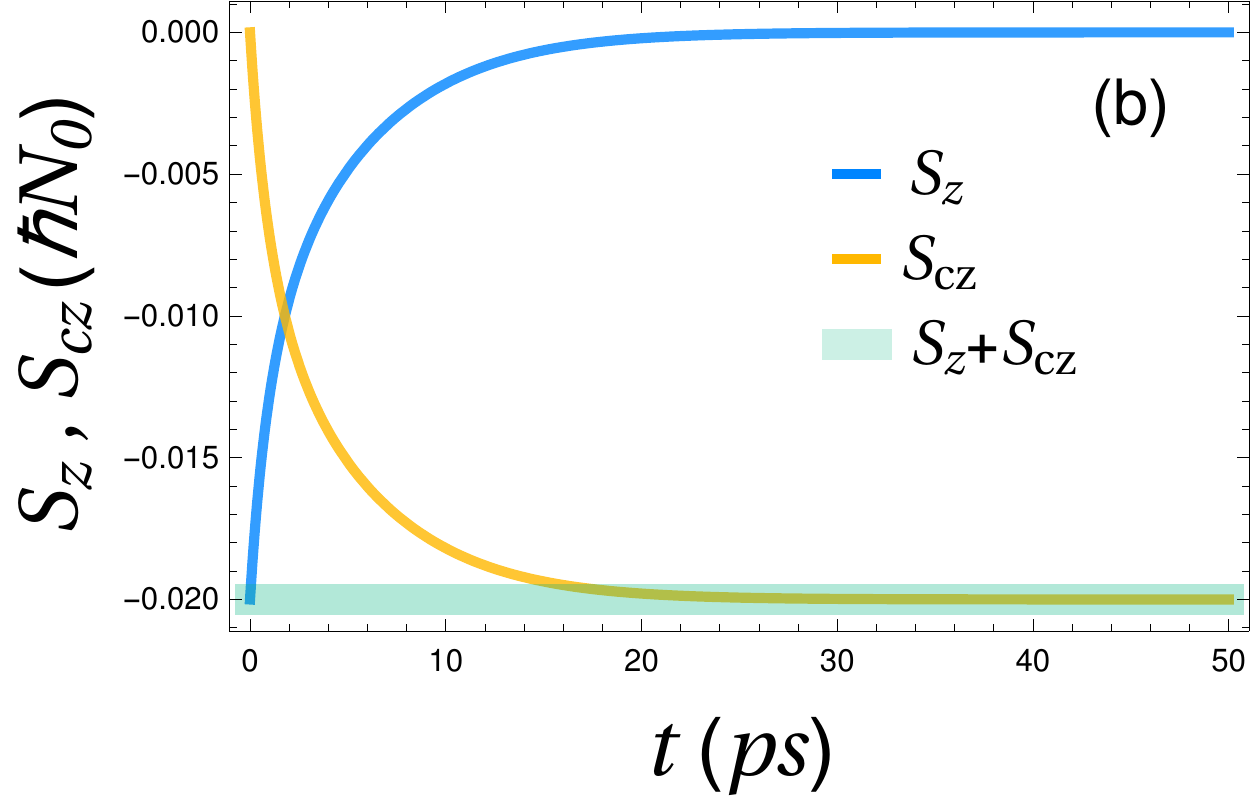}
\includegraphics[width=0.45\textwidth,keepaspectratio=true]{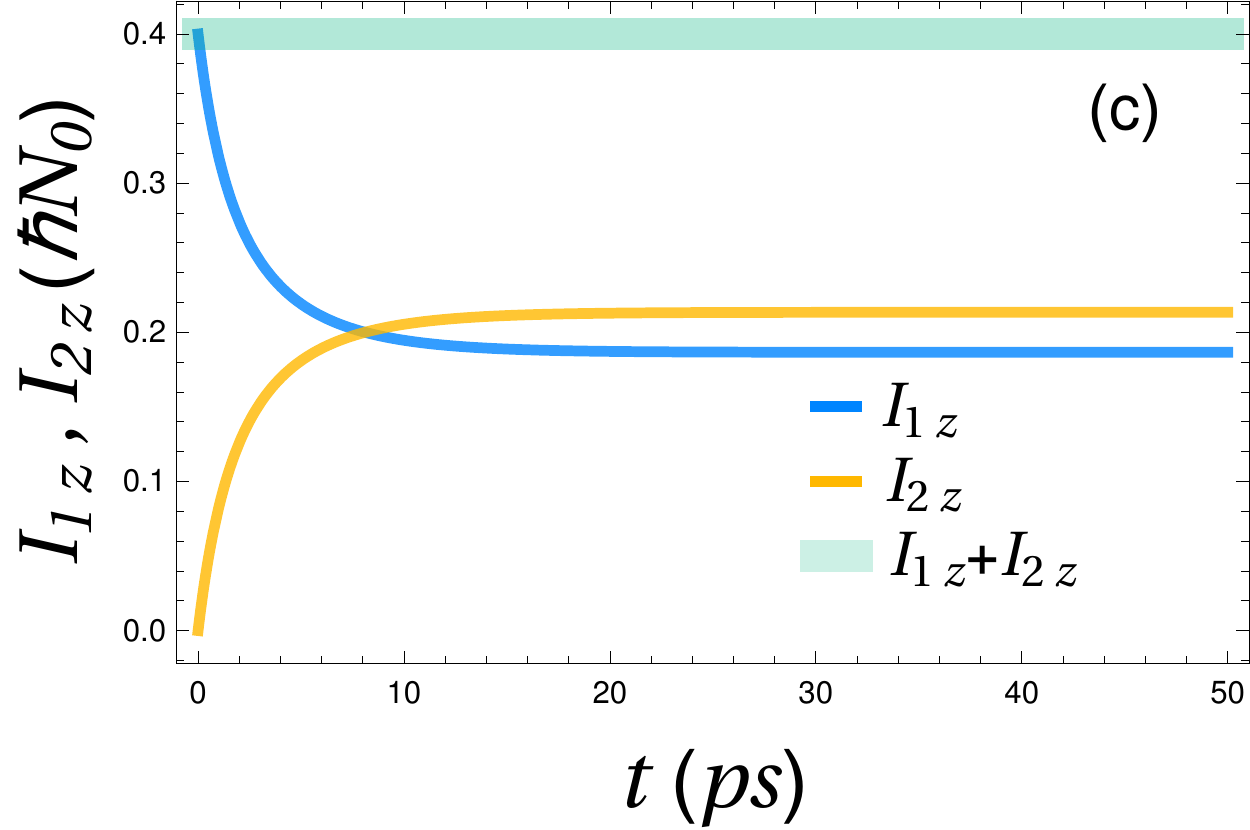}
\includegraphics[width=0.45\textwidth,keepaspectratio=true]{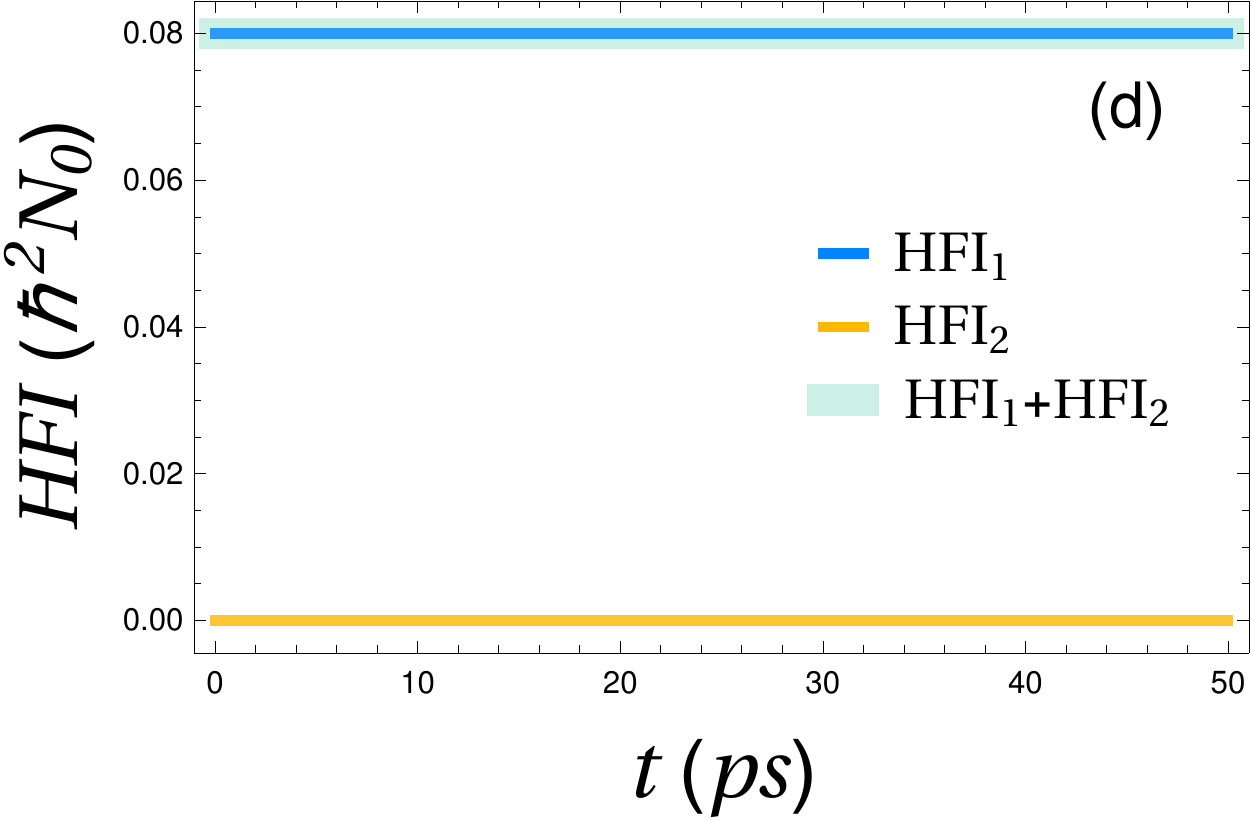}
\caption{(a) Number of singly and doubly occupied centers $N_1$
and $N_2$ as a function of time.
(b) Electronic spin of singly and doubly occupied centers $S_z$
and $S_{cz}$ as a function of time.
(c) Nuclear spin of singly and doubly occupied centers $I_{1z}$
and $I_{2z}$ as a function of time.
(d) Hyperfine interaction as a function of time for
singly and doubly occupied centers. The blue and orange solid lines
correspond to singly and doubly occupied centers respectively.
The thick green line is the sum of singly and doubly occupied centers.
The calculation was performed only taking into account
the SDR dissipator of Eq. (\ref{eq:dissdrnew}) for the following
initial populations
$n_{-1/2}=0.22 N_0$, $n_{-1/2}=0.18 N_0$, $N_{-1/2,-1/2}$
$=N_{-1/2,1/2}$  $=N_{1/2,3/2}=0.2$,
$N_{-1/2,-3/2}=0.08$, $N_{-1/2,3/2}=0.1$, $N_{1/2,-3/2}=0.02$,
$=N_{1/2,1/2}=0.1$.}
\label{fig:figure3}
\end{figure*}

At first glance,
it would seem that the conditions of charge conservation
(\ref{eq:conscharge}),
center conservation
(\ref{eq:constcenters}),
electronic spin conservation
(\ref{eq:conselspin}) and
nuclear spin structure conservation
(\ref{eq:nucspinstrucons})
would suffice to completely define the spin selective capture
of a CB electron. A closer inspection reveals that this
is far from true.
There are 19 elements of the
singly charged center density matrix  $\av{U}_{k,j,i}$
encompassed by the conservation conditions:
population $N^1=8\av{U}_{0,0,0}$,
electronic spin $\av{S}_{ck}=4\av{U_{k,0,0}}$ ($k=1,2,3$)
and nuclear spin structure {$\av{U}_{0,j,i}$}
($i,j=0,1,2,3$ $i\ne 0 \lor j\ne 0$).
Thus, we have overlooked the 45
density matrix coefficients $\av{U}_{k,j,i}$
where $k\ne 0 \land (j\ne 0 \lor i\ne 0)$.
These are in fact associated with the electron-nucleus spin
correlation. Conditions
(\ref{eq:constcenters}),
(\ref{eq:conselspin}) and
(\ref{eq:nucspinstrucons}) in all cases
entail the balance between two reservoirs.
In Eqs. (\ref{eq:constcenters}) and (\ref{eq:nucspinstrucons})
the exchange occurs between
singly and doubly occupied centers
to maintain populations and nuclear spins constant.
The overall electron spin is kept unaltered
through the exchange between CB electrons and
center bounded electrons in
Eq. (\ref{eq:conselspin}).
In contrast, electron-nucleus spin correlation
in Eq. (\ref{eq:ratesdr2})
merely build up in singly occupied centers
due to spin selective recombination.
The correlation unbalance stems from
Eqs. (\ref{eq:ratesdr1})-(\ref{eq:ratesdr3})
where the recombination rate is
modulated by the population in each nuclear state.
Electrons are more likely to recombine
to highly populated nuclear states
consequently distorting
the electron-nucleus spin correlation.
It becomes obvious that the rate equations
(\ref{eq:ratesdrtens1})-(\ref{eq:ratesdrtens3}),
but most particularly (\ref{eq:ratesdr2}),
are beset with a problem when one comes to
realize the HFI is precisely one of the many possible
electron-nucleus spin correlations.
As electrons recombine the
HFI is altered, among other correlations,
and the ratio between the Zeeman interaction
and the HFI is artificially overturned.
It is hardly surprising that the width,
that strongly depends on this ratio,
is incorrectly reproduced by the model.
This fact can be verified in Fig. (\ref{fig:figure2})
where the HFI is plotted as a function of time.
Evidently the HFI is zero for doubly occupied traps
(solid orange line)
since electrons are forming a singlet state that
has vanishing correlation with the nuclear spin.
In contrast, the HFI for singly occupied centers
varies with time (blue solid line), therefore,
the total correlation can not be constant (thick green line
superimposed to the solid blue line).

\begin{figure*}
\includegraphics[width=0.45\textwidth,keepaspectratio=true]{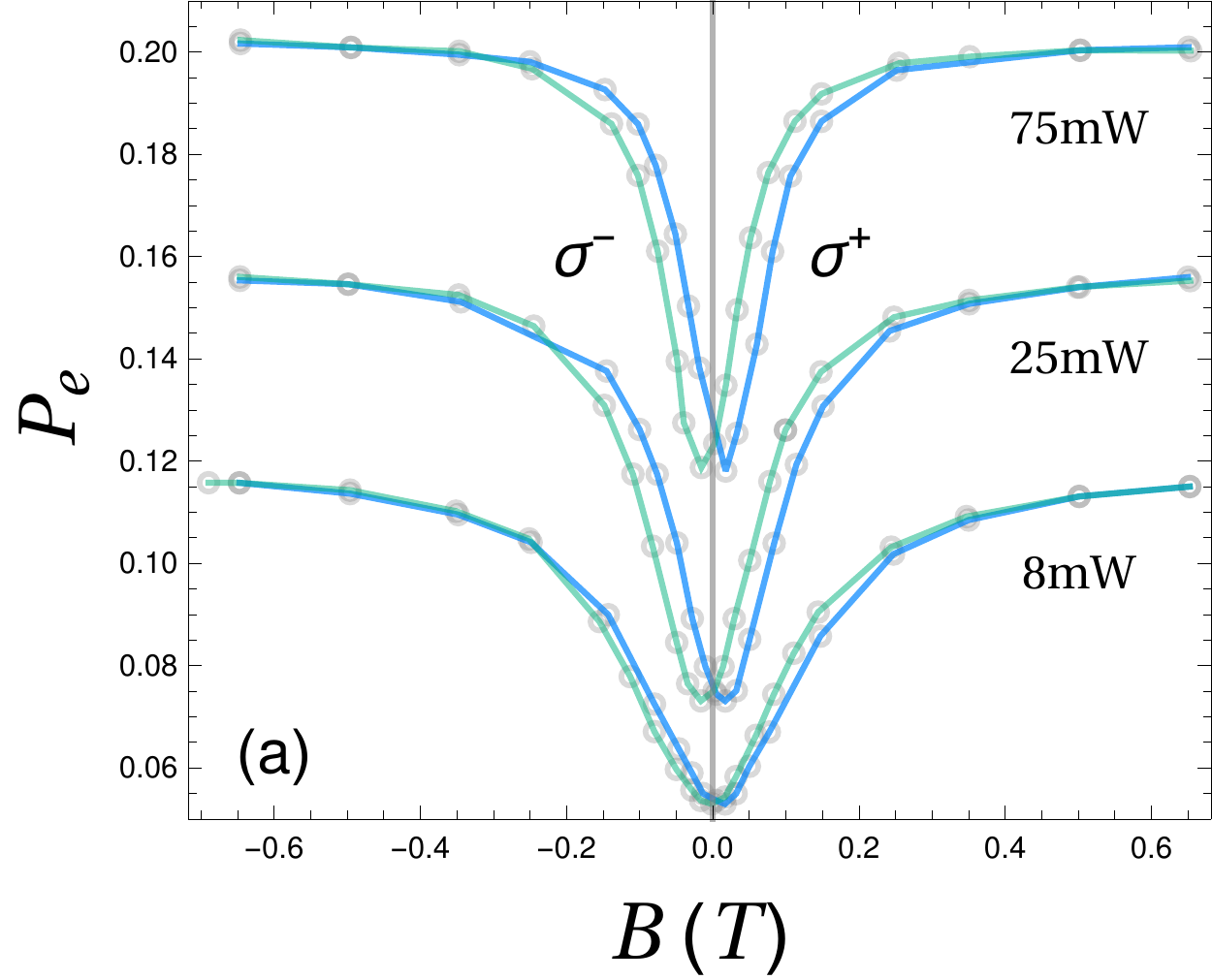}
\includegraphics[width=0.45\textwidth,keepaspectratio=true]{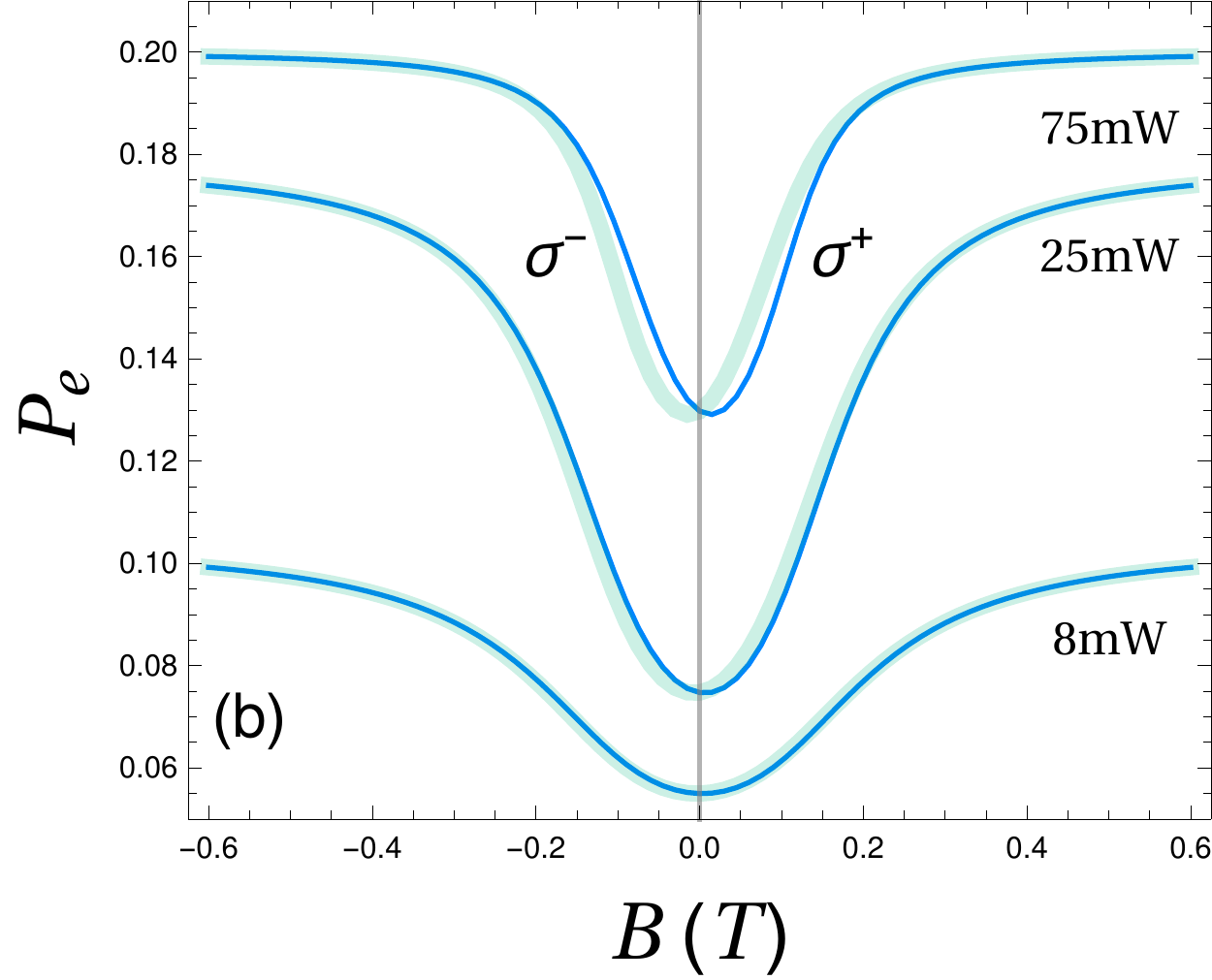}
\caption{(a) Experimental and (b) theroetical results
for the degree of circular polarization as a function
of the Faraday configuration magnetic field $B_z$.
In panel (a) the gray circles indicate the experimental points and
the solid lines are a guide to the eye.
Both panels show the curves corresponding to the
photoluminescence degree of circular polarization
under right circularly polarized excitation ($\sigma^+$, blue solid lines) 
and left circularly polarized excitation ($\sigma^-$, green solid lines).}
\label{fig:figure4}
\end{figure*}

\begin{figure}
\includegraphics[width=0.46\textwidth,keepaspectratio=true]{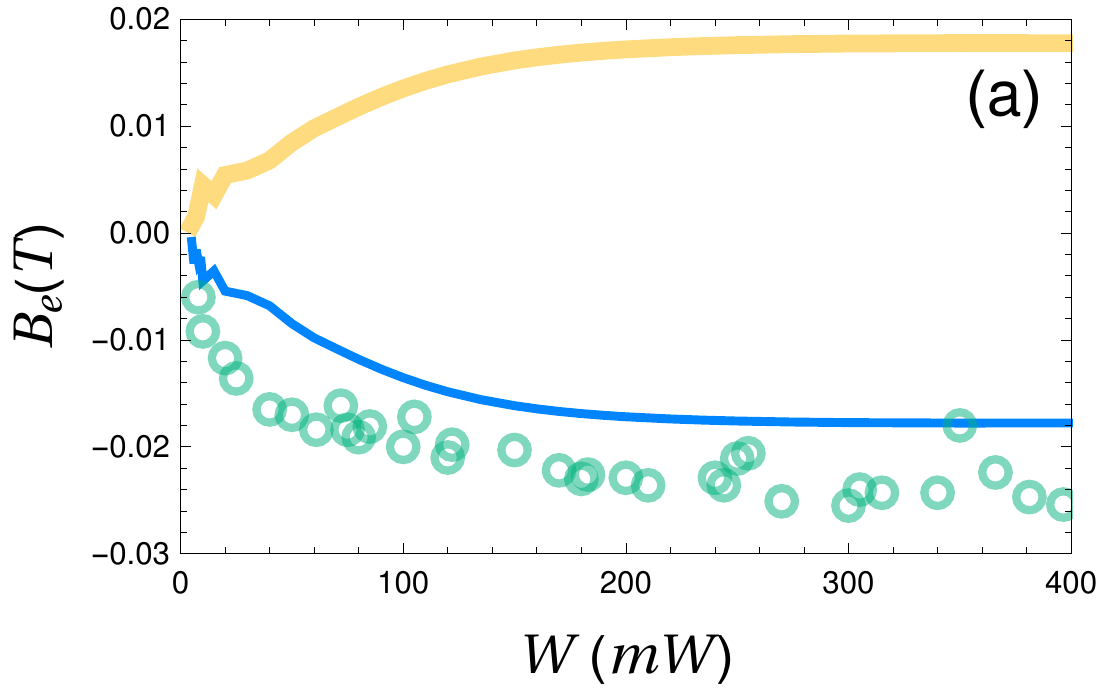}
\includegraphics[width=0.46\textwidth,keepaspectratio=true]{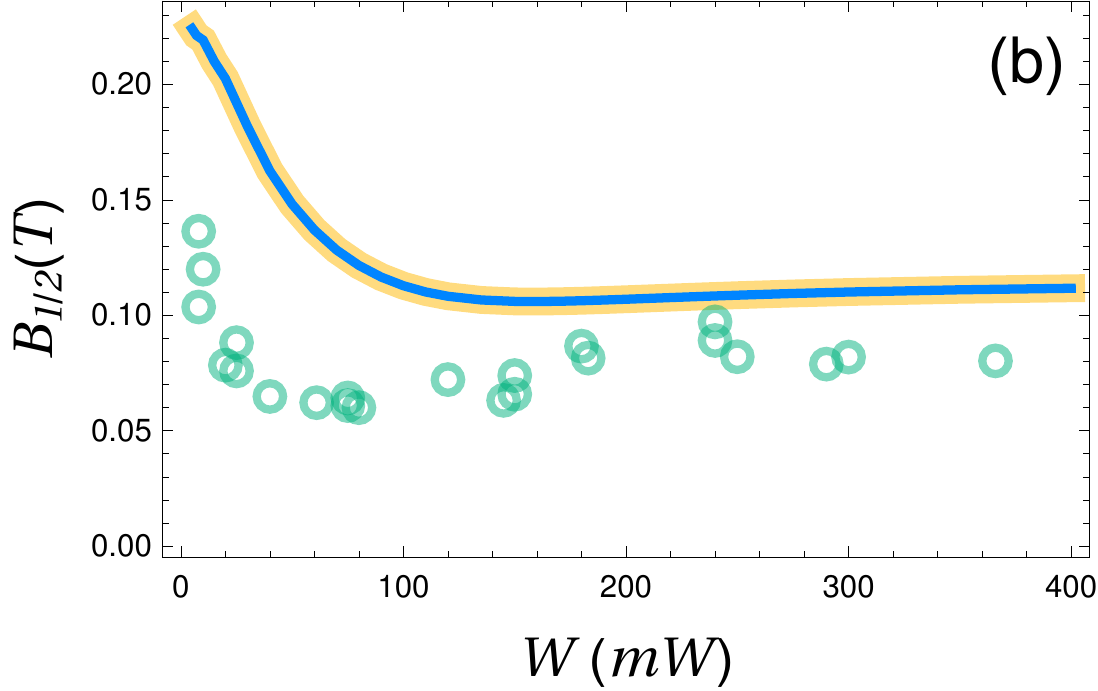}
\includegraphics[width=0.46\textwidth,keepaspectratio=true]{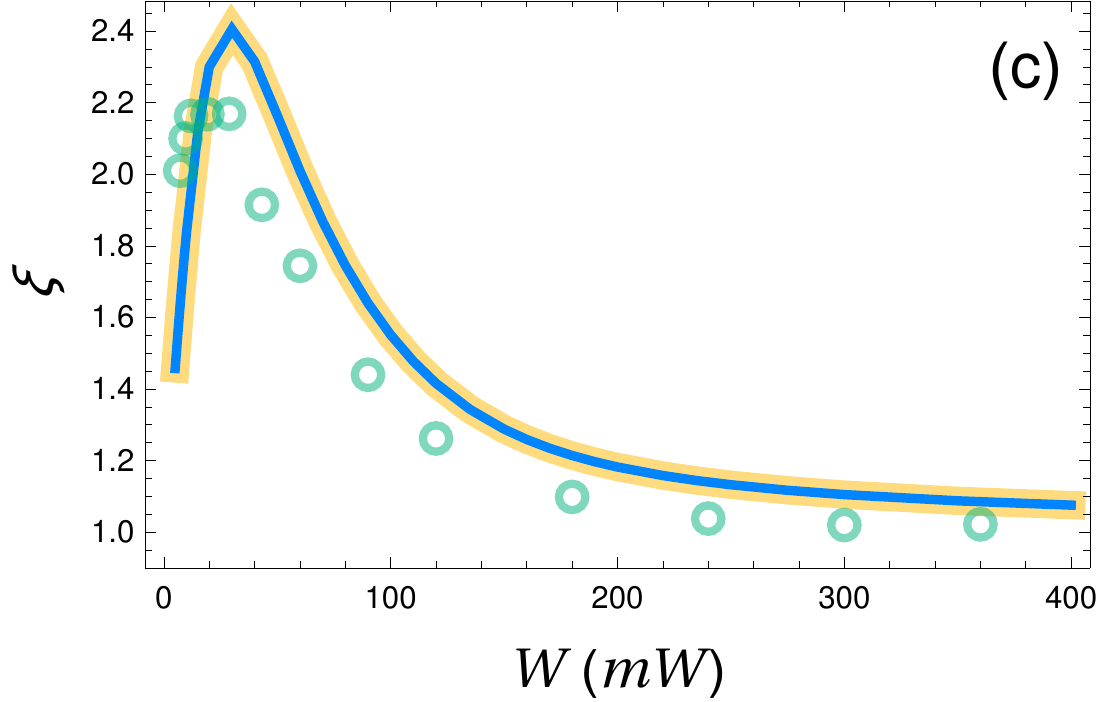}
\caption{(a) Effective magnetic field $B_{\mathrm{e}}$ as a function
of the illumination power.
(b) Half width $B_{1/2}$ as a function of illumination power.
(d) $\xi$ as a function of the illumination power. The solid blue
and orange lines represent the theoretical calculations for right
and left circularly
polarized light respectively. The green circles correspond to
the experimental results.}
\label{fig:figure5}
\end{figure}

\begin{figure}
\includegraphics[width=0.46\textwidth,keepaspectratio=true]{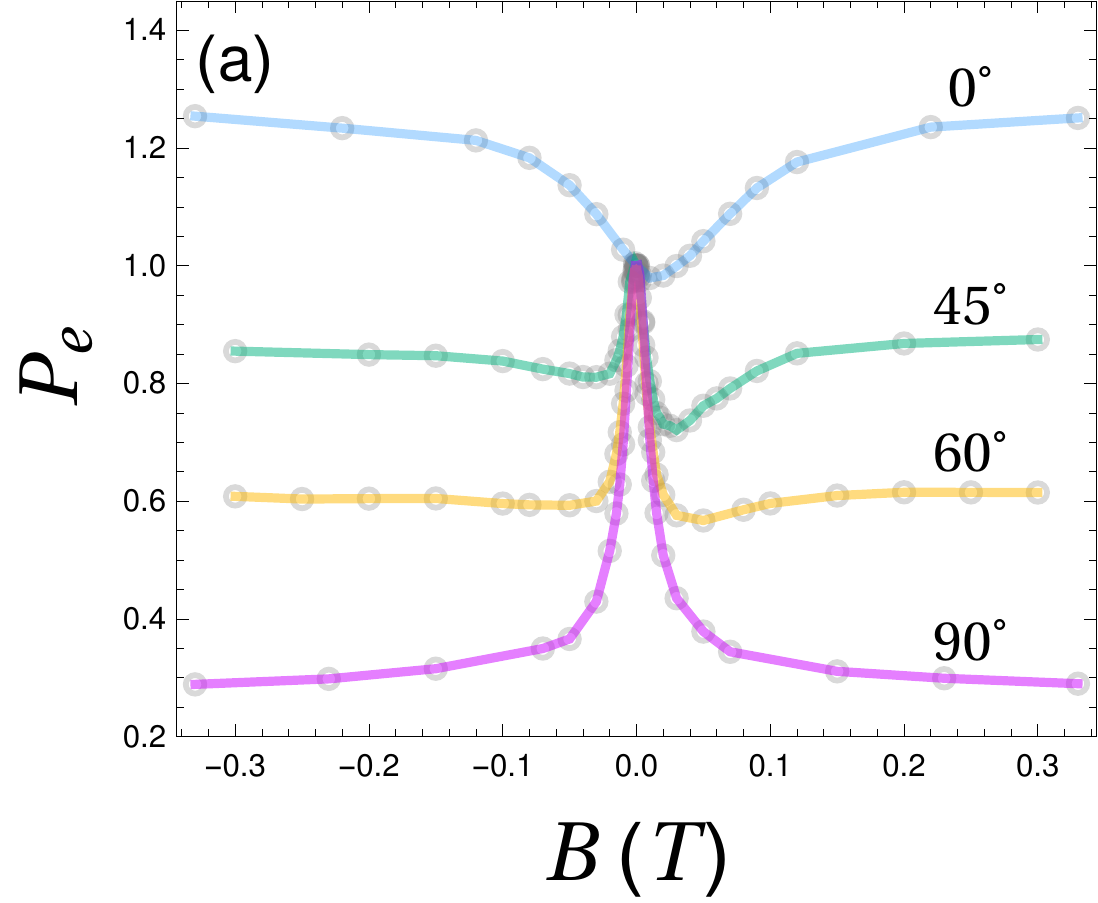}
\includegraphics[width=0.46\textwidth,keepaspectratio=true]{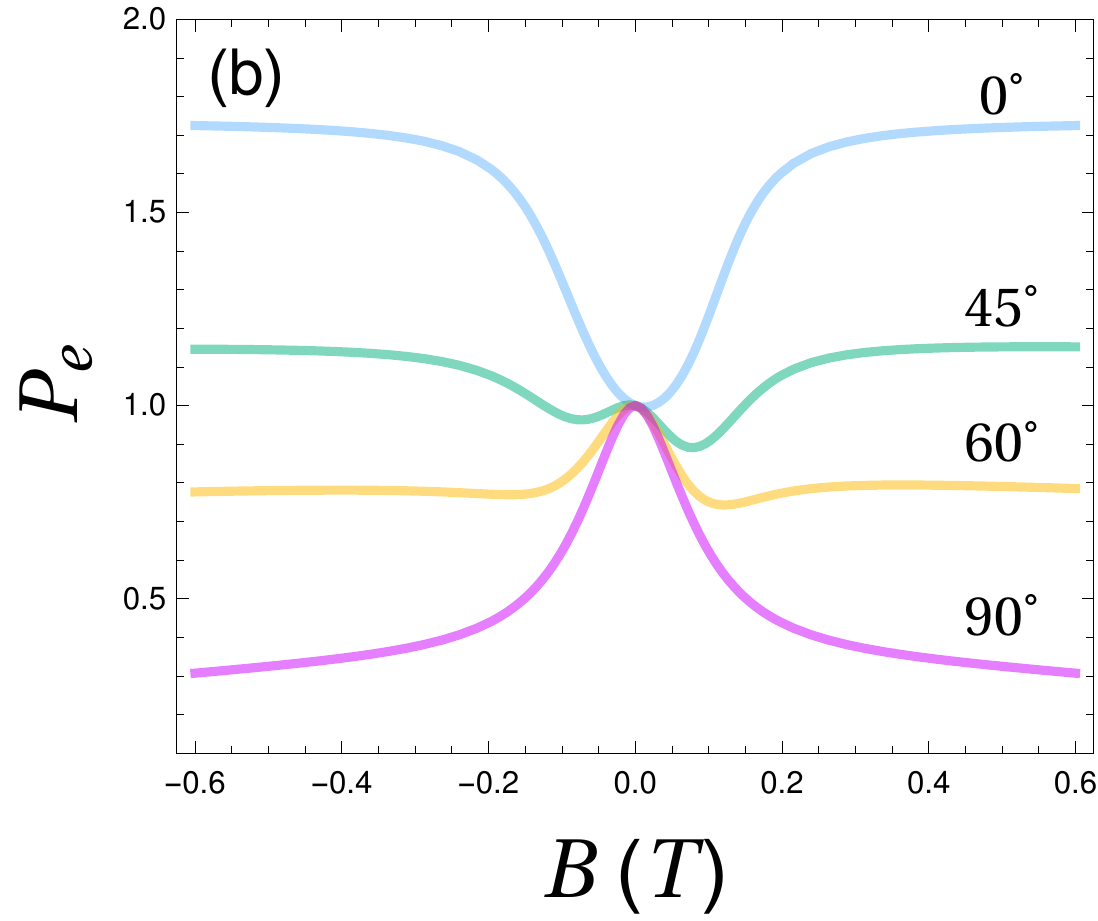}
\caption{(a) Experimental and (b) theoretical results for
degree of circular polarization as a function of an oblique
magnetic field for diverse magnetic field orientations. The incidence
line is perpendicular to the sample and the angles are
measured with respect this orientation.}
\label{fig:figure6}
\end{figure}

The simplest approach to solve this problem
is simply setting to zero the group of equations
that alter the electron-nucleus correlations
without tampering with the spin dependent recombination.
In other words,  we have to
replace Eq. (\ref{eq:ratesdrtens2}) by
\begin{equation}
    \frac{d}{dt}\av{U}_{k,j,i} =
 -c_n\mu_{k,j,i}\sum_{k^{\prime},k^{\prime\prime}=0}^3
  \av{S}_{k^{\prime}}
  Q_{k,k^{\prime},k^{\prime\prime}}
  \av{U}_{k^{\prime\prime},j,i}\, ,\label{eq:newratesdrtens2}
\end{equation}
where
\begin{equation}
\mu_{k,j,i}=\begin{cases}
1\,\,\,\, , k=0 \lor (k\ne 0 \land j=0\land i=0) \,\, ,\\
0\,\,\,\, , k\ne 0 \land (j\ne 0 \lor i\ne 0)\,\, . \\
\end{cases}
\end{equation}
With this modification the $\mathcal{D}_{SDR}$ dissipator
takes the form
\begin{multline}
\mathcal{D}_{SDR}=
  -2\sum_{k=0}^3
  \left(4c_n\sum_{k^{\prime},k^{\prime\prime}=0}^3
  \av{S}_{k^{\prime}}
  Q^{\top}_{k,k^{\prime},k^{\prime\prime}}
  \av{U}_{k^{\prime\prime},j,i}\right)
  S_{k}\\
 -8\sum_{k,j,i=0}^3\mu_{k,j,i}
 \left(c_n\sum_{k^{\prime},k^{\prime\prime}=0}^3
  \av{S}_{k^{\prime}}
  Q_{k,k^{\prime},k^{\prime\prime}}
  \av{U}_{k^{\prime\prime},j,i}\right)
  U_{k,j,i}\\
 +4\sum_{j,i=0}^3
  \left(2c_n\sum_{k^{\prime},k^{\prime\prime}=0}^3
  \av{S}_{k^{\prime}}
  Q^{\top}_{0,k^{\prime},k^{\prime\prime}}
  \av{U}_{k^{\prime\prime},j,i}\right)
  V_{j,i},\label{eq:dissdrnew}
\end{multline}
Working back the population rate equations
we find that the only change occurs in
in Eq. (\ref{eq:ratesdr2}) that now takes
the form
\begin{multline}
\frac{d}{dt}\av{N}^1_{\alpha,\beta} =
  -\frac{c_n}{8}\sum_{\beta^{\prime}=-3/2}^{3/2}
  \left(\av{n}_{-\alpha}\av{N}^1_{\alpha,\beta^{\prime}}
  -\av{n}_{\alpha}\av{N}^1_{-\alpha,\beta^{\prime}}\right)
  \\
  -\frac{c_n}{2}\sum_{\alpha^{\prime}=-1/2}^{1/2}
 \av{n}_{-\alpha^{\prime}}\av{N}^1_{\alpha^{\prime},\beta}.
  \label{eq:newratesdr2}
\end{multline}
Because the structures of (\ref{eq:ratesdr1}) as well as
that of (\ref{eq:ratesdrtens2}) for  $k\ne 0 \land j=0\land i=0$
were not modified, the spin selective capture of electrons
remains unaffected keeping the essential properties
of the two-charge-state model.
The first term in the right hand side of the previous
equation is antisymmetric in the electronic spin index
and hence, gives rise to the spin dependent capture of electrons.
By itself, it does not change the overall population
of singly occupied centers but instead shifts the occupation
number to the bound electron spin states that have the same
spin orientation as the majority of the CB electrons.
The second term, on the other hand, is symmetric in the
electronic spin index and therefore is responsible of the
population reduction following the capture of an electron.
These terms have such symmetries that
all the nuclear states are depopulated at the same speed
keeping the electron-nucleus spin correlation constant.
It should be stressed that even though (\ref{eq:dissdrnew})
gives very good results, it
is not the only possible dissipator that one
can think of in order to guarantee constant electron
and nuclear spin correlations. In Appendix \ref{ap:appendixa}
we present the most general form that this dissipator
has to take to ensure correlation conservation.

The new dissipator (\ref{eq:dissdrnew}) in any of its forms is not
without problems. From the structure of (\ref{eq:newratesdr2}),
it is clear that it
can not ensure the positive definitness of the density matrix
because, contrary to (\ref{eq:ratesdr2}),
the recombination rate might not become zero
once the whole population of a given state
is depleted. In simple terms, if the nuclear
sates have different populations and depopulate
at the same velocity, at least one of them
is going to run out of electrons before the others.
It is therefore impossible to simultaneously endow the
dissipator with positive definitness and constant
electron-nucleus correlation.
To circumvent this difficulty we have verified in
every calculation the positive definitness of the
density matrix, i.e., that populations
are in fact positive.
\blue{$comment:$ how this is achieved? Is it pure chance that in this case the population stay positive or there is something built-in to prevent it?}

To illustrate the dynamics of these conservation principles
under the action of the dissipator (\ref{eq:dissdrnew}),
in Fig. \ref{fig:figure3}
we show the time-dependence of the number of centers,
electronic spins, nuclear spins and electron-nucleus correlations
as functions of time for initially non homogeneously occupied
electronic and nuclear spin states.
As indicated by the orange solid lines, we observe that
overall values of these quantities are preserved over
a spin dependent recombination process.
Particular attention should be payed to Fig. \ref{fig:figure3}d
where it is shown that, unlike (\ref{eq:dissdr}),
the dissipator (\ref{eq:dissdrnew}) preserves the average of the HFI.
It should be noted that in this figure we only plot
one of the 45 possible electron-nucleus correlations. The
remaining 44 correlations are also preserved although
they are not shown here.

\section{Results and discussion}\label{sec:resanddisc}
Now we want to study in more detail the effect of
the new SDR dissipator $\mathcal{D}_{SDR}$ in Eq.(\ref{eq:dissdrnew})
in measurable quantities as the DCP
$P_{\mathrm{e}}(B_z)$
and the PL intensity $J(B_z)$.
In particular, we intend to verify if the new structure
of $\mathcal{D}_{SDR}$ is capable of capturing the 
main experimental features of the
shift $B_{\mathrm{e}}$
and width $B_{1/2}$ as functions of the illumination power.
Both, the DCP and the PL intensity yield similar curves
for the shift and width as functions of power, therefore, we only
concentrate in the behaviour of the $B_{\mathrm{e}}$ and $B_{1/2}$
that stem from $P_{\mathrm{e}}(B_z)$.
To further examine the performance of the model in
reproducing experimental results we contrast
the measurements of the DCP in oblique magnetic fields \cite{Ivchenko2016}
This test is of particular interest because
under a titled magnetic field
$P_{\mathrm{e}}(B_z)$ strongly depends on the width
and the shift. Moreover, previous models based on
simpler nuclear structures with spin $1/2$ \cite{Ivchenko2016}
fail to reproduce some of the features.
In addition, the quantities that were correctly reproduced
by the old SDR dissipator $\mathcal{D}_{SDR}$ in Eq.(\ref{eq:dissdr})
are expected to maintain their previous trends.

Using the selection rules of GaAs \cite{meier2012optical}
it can easily be proven that the DCP is related
to the degree of spin polarization of
CB electrons as
\begin{equation}
    P_{\mathrm{e}}=\frac{P_i}{3}\frac{\av{S}_z(t)}{2\av{n}(t)},
\end{equation}
where $P_i$ is a phenomenological factor\cite{PhysRevB.85.035205}.
To extract the expectation values of
CB electrons spin and population $\av{S}_z(t)$
and $\av{n}(t)$ we introduce the operators
$S_z$ and $n$ into Eq. (\ref{eq:expectation})
along with the expansion of the density matrix $\rho(t)$
(\ref{eq:densmat}).
This procedure casts
$\av{S}_z(t)$
and $\av{n}(t)$, and any other observable,
into the convenient form of a function of the quantum-statistical
averages $\av{\lambda}_q(t)$.
This way of expressing the quantum-statistical averages
is specially suitable to make calculations
in an integrated manner with the
master equation of the density matrix.
It only remains to transform
the master equation (\ref{eq:master}),
a set of differential equation for the elements of $\rho(t)$,
into a series of differential equations
for the quantum-statistical averages $\av{\lambda}_q$.
This is easily achieved through multiplying the master equation by
the operator $\lambda_q$ and taking the trace \cite{PhysRevB.95.195204}
\begin{multline}
  \dot{\av{\lambda}}_q =\frac{i}{\hbar}\Tr\left[\left[H,\lambda_q\right]\rho(t)\right]
  +\Tr\left[\mathcal{D}\lambda_q\right]\\
  =F_q(\lambda_1,\lambda_2,\dots,\lambda_d,t).\label{eq:newrateeq}
\end{multline}
Replacing the density matrix
with the expansion (\ref{eq:densmat})
we are led to the result
\begin{multline}
    F_q(\lambda_1,\lambda_2,\dots,\lambda_d,t)\\
      =\frac{i}{\hbar}\sum_{q^\prime=1}^{d}
      \frac{\av{\lambda}_{q^\prime}}{\Tr[\lambda_{q^\prime}^2]}
     \Tr\left[H\left[\lambda_q,\lambda_{q^\prime}\right]\right]
     +\Tr[\mathcal{D}\lambda_q].
\end{multline}
The $d=85$ differential equations that arise from
(\ref{eq:newrateeq}) are in fact
the new rate equations that generalize
the two-charge-state model \cite{PhysRevB.95.195204}.
Any quantum-statistical average is calculated
by numerically solving
the system of ordinary
differential equations (\ref{eq:newrateeq}),
allowing it to reach steady state conditions
and plugging the solution into Eq. (\ref{eq:expectation}).

In particular, here we contrast three specific
theoretical results with its experimental counterparts:
$B_{\mathrm{e}}$, $B_{1/2}$ and $\xi$ as functions of
the illumination power
where
\begin{equation}
\xi=P_{\mathrm{e}}(\infty)/P_{\mathrm{e}}(0).
\end{equation}
In the previous equation $P_{\mathrm{e}}(0)$ is the DCP
at zero longitudinal magnetic field
and $P_{\mathrm{e}}(\infty)$ is the DCP at infinite
longitudinal magnetic field. At infinite magnetic
field, the Zeeman interaction overwhelms
the HFI decoupling the electronic and nuclear spins
in Ga$^{2+}$ centers. In these conditions the system
behaves as if there was no HFI and could in principle
be described solely by the two-charged state model.
Hence $\xi$ parametrizes the degree of participation
of the HFI in the SDR process.

To start,
Fig. \ref{fig:figure4} shows a comparison of the
experimental and theoretical results of $P_{\mathrm{e}}$
as a function of the Faraday configuration magnetic field
for various illumination powers.
The parameters that best fit the experimental data
are the following: $N_0=2.24\times 10^{15}$cm$^{-3}$,
$\tau^*=5$ps, $\tau_h=30$ps, $\tau_s=120$ps,
$\tau_{sc}=2100$ps, $\tau_{n1}=2100$ps, $\tau_{n2}=5$ps,
$G_0=2.0\times 10^{23}$mW$^{-1}$s$^{-1}$cm$^{-3}$, $P=0.18$,
$P_i=0.3$, $g=1$, $g_c=1.7$ and $A=0.0690$cm$^{-1}$.
In contrast with the results shown in
Fig. \ref{fig:figure1} or
in Ref. [\onlinecite{PhysRevB.95.195204}],
the effective magnetic field and the half width
saturate at approximately $150$mW.
This is confirmed in Fig. \ref{fig:figure5} where
$B_{\mathrm{e}}$, $B_{1/2}$ and $\xi$ are plotted
as functions of the illumination power $W$.
The effective magnetic field
$B_{\mathrm{e}}$ and the mean width $B_{1/2}$ have been extracted
from the $P_{\mathrm{e}}(B_z)$ curves by means of the
{ golden section search} algorithm with a tolerance of $1\mu$T.
$P_{\mathrm{e}}(0)$ and $P_{\mathrm{e}}(\infty)$, necessary
to calculate $\xi$, are byproducts of the algorithm
used to determine $B_{1/2}$.
The experimentally determined effective magnetic
field $B_{\mathrm{e}}$ in panel (a) of Fig. \ref{fig:figure5}
(green circles) starts at $0$T and decreases steadily with power
down to $-25$mT at approximately $150$mW.
Similarly, the theoretical results for $B_{\mathrm{e}}$
(solid blue lines) yield a monotone decreasing function of power
that saturates at approximately $-15$mT at a power of $150$mW
for left circularly polarized light ($P=0.18$).
Under right circularly polarized excitation ($P=-0.18$)
we obtain the opposite result:
the effective magnetic field increases with power
until it reaches a point of saturation at $+0.15$mT
for a power of $150$mW (solid orange line).
Even though there is a difference of $10$mT between
the experimental and theoretical
saturation effective magnetic fields, both trends are qualitatively
comparable. It is unlikely that this difference is due to
the parameter choice since a vast number of
parameter combinations yield $\vert B_{\mathrm{e}}\vert=\pm 15$mT
as the maximum obtainable
value for the effective magnetic field. For instance, following a similar
procedure as the one described in Ref. [\onlinecite{PhysRevB.95.195204}],
the $B_{\mathrm{e}}$ isolines as a function of the nuclear
spin relaxation times $\tau_{n1}$ and $\tau_{n2}$ give a maximum
value of approximately $15$mT for $\tau_{n1}=2100$ps and $\tau_{n2}<5$ps.
It is more probable that this difference is due
to the particular choice of dissipator made in (\ref{eq:dissdrnew}).
As we mentioned above and in appendix \ref{ap:appendixa}, the electron nucleus
spin correlation conservation condition alone is not
sufficient to completely define the SDR dissipator.
The behaviour of the calculated mean width $B_{1/2}$ as a
function of power (solid blue line), shown in
Fig. \ref{fig:figure5} (b), exhibits a very good
qualitative and quantitative
agreement with the experimental data (green circles).
At low powers we observe the largest discrepancy between experimental
and theoretical results: the theoretical values of $\xi$ underestimate
by a few tens of mT the experimental ones.
Both trends are quite similar: $B_{1/2}$ decreases monotonically
until it saturates at approximately $200$mT.
The curve displayed by the theoretical values of $\xi$ as a function
of illumination power seen in Fig. \ref{fig:figure5} (c) is also in very
good agreement with the experimental data.
This curve had been correctly reproduced by previous models \cite{PhysRevB.95.195204}.
This is to be expected since $\xi$ solely depends on two
extreme situations: the first, when the HFI is the dominant
interaction ($B_z=0$) and the second, when
the Zeeman energy prevails over the HFI ($B_z\rightarrow \infty$).
It does not depend however on \blue{which} value of the magnetic field $B_z$
the HFI becomes irrelevant and \blue{on }how steep this transition \blue{is}.
Thus, the agreement between the experimental and theoretical
results for $\xi$ is an indication that the new SDR dissipator
captures the correct behaviour of $P_{\mathrm{e}}$
at both ends, $B_z=0$ and $B_z\rightarrow \infty$, but at the same time
modifies the magnetic field value at which the
electrons and nuclei in Ga$^{2+}$ centers
transition from having a strong hyperfine coupling
to being decoupled.

To \blue{conclude} our discussion and further test
the capabilities of the new SDR dissipator, we move on
to the experimental results of the DCP $P_{\mathrm{e}}$
in an oblique magnetic field \cite{Ivchenko2016}.
Figure \ref{fig:figure6} (a) shows the experimental
data of the $P_{\mathrm{e}}$ as
a function of the magnitude of a tilted magnetic field
for various magnetic field orientations.
Once again, here we attain a very good agreement
with the theoretical calculations plotted
in Fig. \ref{fig:figure6} (b).
At $90^\circ$ (solid purple line)
the $P_{\mathrm{e}}$ exhibits the typical Lorentzian curve
of the Hanle effect
observed in Voigt configuration \cite{KALEVICH20094929}.
In contrast, at $0^\circ$ (solid blue line) we observe the characteristic
inverted Lorentzian curve
corresponding to the amplification of the spin filtering
effect \cite{PhysRevB.85.035205}.
The intermediate angles ($45^\circ$ and $60^\circ$)
yield a superposition of both functions \cite{Ivchenko2016}:
the upward Hanle effect Lorentzian and the
downward Lorentzian corresponding to the amplification
of the spin filtering effect.

\section{Conclusions}

We have sistematically investigated the
consequences of the bogus 
electron-nucleus spin correlations
in the spin dependent capture
of electrons in Ga$^{2+}$ paramagnetic centers.
These were inadvertently incorporated
in most models through the bimolecular-like
terms that account for the mechanism
of spin dependent recombination of
CB electrons in Ga$^{2+}$ defects.
We have shown that the electron-nucleus
spin correlations are responsible of the
pronounced differences between the
experimental and theoretical findings
on the effective magnetic field $B_{\mathrm{e}}$
and width $B_{1/2}$
as functions of the illumination power.
The general form of an alternative spin dependent capture
mechanism that preserves electron-nucleus
spin correlation has been proposed
and thoroughly tested.
This mechanism, embedded in \blue{the} master equation
for GaAsN in the form of a dissipator,
yields very good agreement between
theoretical and \blue{the} experimental observations.
\blue{particular}, very good accordance is observed
with experimental data concerning $B_{\mathrm{e}}$,
$B_{1/2}$ and $P_{\mathrm{e}}(B_z)$.

\section{Acknowledgements}
We acknowledge funding from LIA CNRS-Ioffe RAS ILNACS.  L.A.B. and E.L.I. thanks the Russian Foundation for Basic Research (Grants No. 17-02-00383 and No. 17-52-16020). V.K.K. acknowledges the financial support of the Government of Russia (Project No. 14.Z50.31.0021).  A.K. gratefully 
appreciates the financial support of Departamento de Ciencias B\`asicas UAM-AÓ grant numbers 2232214 
and 2232215. V.G.I.S and J.C.S.S. acknowledge the total support from 
DGAPA-UNAM fellowship. X.M. also thanks Institut Universitaire de France.
This work was supported by Programme Investissements d'Avenir under the program ANR-11-IDEX-0002-02, reference ANR-10-\blue{LABX}-0037-NEXT.

\appendix
\section{}\label{ap:appendixa}
In order to avoid adding electron-nucleus spin
correlations, 
the most general form of the SDR rate equations
(\ref{eq:ratesdrtens1})-(\ref{eq:ratesdrtens3})
should be
\begin{eqnarray}
\frac{d}{dt}\av{S}_k = \eta_k,\label{eq:generalrate1} \\
\frac{d}{dt}\av{U}_{k,0,0} = -\frac{\eta_k}{4},&&
  \,\,\, k=0,1,2,3, \label{eq:generalrate2} \\
\frac{d}{dt}\av{U}_{0,j,i} = \varphi_{j,i}, &&
  \,\,\, j,i=0,1,2,3, / \{i=j=0\}, \label{eq:generalrate3}\\
\frac{d}{dt}\av{U}_{k,j,i} = 0, && \,\,\, k=1,2,3,
\nonumber \\
 && \,\,\, i,j=0,1,2,3, / \{j=i=0\},\label{eq:generalrate4}\\
\frac{d}{dt}\av{V}_{j,i} = -2 \varphi_{j,i}, &&
  \,\,\, j,i=0,1,2,3, , \label{eq:generalrate5}
\end{eqnarray}
where $\eta_k$ and $\varphi_{j,i}$ are arbitrary
generation rate terms.
Equation (\ref{eq:generalrate4}) guarantees that
no extra electron-nucleus spin correlation is added during
the spin selective capture of an electron.
The generation rates in Eqs. (\ref{eq:generalrate1}) and (\ref{eq:generalrate2})
are balanced to preserve electronic spin during the recombination
process.
Similarly, the generation rates of Eqs.
(\ref{eq:generalrate3}) and (\ref{eq:generalrate5})
compensate to maintain constant nuclear spin and total
center population.
The following supplementary constraint regarding charge conservation
must be added to this system of equations
\begin{equation}
  4\varphi_{0,0}-2\eta_0=0.\label{eq:chargeconsap}
\end{equation}
If additionally we assume that the recombination
process must also have
the same structure as the two charged-state model
then, from Eq. (\ref{eq:ratesdrtens1}) we obtain
\begin{equation}
    \eta_k=-4c_n\sum_{k^{\prime},k^{\prime\prime}=0}^3
  \av{S}_{k^{\prime}}
  Q^{\top}_{k,k^{\prime},k^{\prime\prime}}
  \av{U}_{k^{\prime\prime},j,i}.
\end{equation}
Using this equation and (\ref{eq:chargeconsap})
one may also determine $\varphi_{0,0}$.
The remaining recombination rates of the form $\varphi_{j,i}$
for $j,i=0,1,2,3$ except $j=i=0$ are undetermined and
can not be derived from any conservation principle.
In the rate equation (\ref{eq:ratesdrtens3})
we made the obvious choice of setting the remaining
recombination rates to
\begin{equation}
    \varphi_{j,i} = 2c_n\sum_{k^{\prime},k^{\prime\prime}=0}^3
  \av{S}_{k^{\prime}}
  Q^{\top}_{0,k^{\prime},k^{\prime\prime}}
  \av{U}_{k^{\prime\prime},j,i}.
\end{equation}
This, however, is not the only possibility.


%

\end{document}